\documentclass[a4paper,12pt]{article}

\usepackage{jheppub}
\usepackage[T1]{fontenc}
\usepackage{amssymb}
\usepackage{graphicx}
\usepackage{amsmath}
\usepackage{changes}
\usepackage{hyperref}
\usepackage{tensor}
\usepackage{epstopdf}
\usepackage{enumitem}
\usepackage{extarrows}
\usepackage{xcolor}
\usepackage{verbatim}
\usepackage{subfigure}
\usepackage{amsmath}
\usepackage{amsthm}
\usepackage{braket}
\usepackage{subfigure}
\usepackage{tikz}
\newcommand{\td}{\text{d}}
\newcommand{\te}{\text{e}}
\usetikzlibrary{arrows, positioning}
	\definechangesauthor[name={Per cusse}, color=orange]{per}
\begin{document}

\title{Application of Solving Inverse Scattering Problem in Holographic Bulk Reconstruction}%

\author[a]{Bo-Wen Fan,}
\author[a]{Run-Qiu Yang}
\emailAdd{cheidaisuki@tju.edu.cn}
\emailAdd{aqiu@tju.edu.cn}
\affiliation[a]{Center for Joint Quantum Studies and Department of Physics, School of Science, Tianjin University, Yaguan Road 135, Jinnan District, 300350 Tianjin, P.~R.~China}
\abstract{We investigate the problem of bulk metric reconstruction in holography by leveraging the inverse scattering framework applied to boundary two-point correlation functions.  We generalize our previous work of scalar field and show that reconstruction can be achieved using a single operator rather than a pair. We also apply this method into reconstruction of static homogeneous anisotropic black holes and the reconstruction using correlation function of gauge field. In addition, we analyze the method's robustness under measurement noise and propose filtering strategies to improve reconstruction accuracy. This work advances data-driven bulk reconstruction by providing a concrete, experimentally viable pathway to recover spacetime geometry from field-theoretic observables.}

\maketitle
\section{Introduction}
The anti-de Sitter (AdS)/conformal field theory (CFT) correspondence \cite{Maldacena:1997re}, which is also known as holography principle, provides a powerful framework for studying strongly coupled quantum field theories through general relativity. After this theory was proposed, it found widespread applications in fields such as quantum information, condensed matter physics, and high energy physics. One notable example is the GKPW master rule introduced by Gubser, Polyakov, Klebanov, and Witten \cite{Gubser:1998bc,Witten:1998qj}, which establishes that the generating functional of the boundary field theory can be obtained by computing the on-shell action of the bulk gravitational theory. This rule not only quantifies the correspondence between partition functions but also maps bulk fields to boundary operators, enabling direct computation of correlation functions through bulk solutions \cite{Witten:1998zw}.

In most of current studies on holographic principle, problems generally fall into two categories: the "direct problem" of deriving boundary observables from known bulk configurations, and the "inverse problem" of reconstructing unknown bulk structures from boundary data. The latter, termed bulk reconstruction, has become central to both theoretical understanding of spacetime emergence and practical applications in holographic modeling \cite{Baggioli:2019rrs,Harlow:2018fse}. Although we have much information on the path that connects AdS and CFT, from a theoretical perspective, bulk reconstruction is central to understanding the emergence of spacetime in holography and explaining how gravitational degrees of freedom arise from field theory. Practically, it allows for more efficient construction of holographic spacetime and theory by directly using observable data when employing holographic duality to tackle strongly coupled materials. Its significance stems from two perspectives: Theoretically, it illuminates how gravitational degrees of freedom emerge from field theory and the nature of spacetime in holography. Practically, it enables efficient spacetime construction using observable data when applying holographic duality to strongly coupled systems like quantum materials. Several advances have been made in this area. When the conformal dimension $\Delta$ is much larger than the spacetime dimension $d$, boundary two-point correlation functions can be expressed in terms of geodesics \cite{Balasubramanian:1999zv} and used to reconstruct bulk geometry \cite{Hubeny:2012ry}. Another method, known as bulk-cone singularities, uses the relationship between singular two-point functions and bulk null geodesics to recover certain spherically symmetric static bulk geometries \cite{Bilson:2008ab,Hammersley:2006cp}. This approach is applied into the spherically symmetric situations and requires to identify all singularities of two-point functions in boundary spacetime coordinates.

Another well-known reconstruction method is entanglement wedge reconstruction, which uses the boundary entanglement spectrum to reconstruct bulk geometry and local observables \cite{Dong:2016eik,Espindola:2018ozt,Penington:2019npb,Faulkner:2017vdd}. Entanglement wedge reconstruction utilizes boundary entanglement spectra \cite{Dong:2016eik,Penington:2019npb}, and in general situations it  requires complete entanglement data of the boundary theory. Other approaches involve holographic entanglement entropy \cite{Ryu:2006bv,Nishioka:2009un,Jokela:2020auu,Ahn:2024lkh}, Wilson loops \cite{Maldacena:1998im,Rey:1998ik}, tensor networks \cite{Swingle:2009bg,Pastawski:2015qua,Cao:2020uvb,Qi:2013caa,Vidal:2008zz}, conformal symmetry \cite{Nakayama:2015mva,Verlinde:2015qfa,Czech:2016xec,Anand:2017dav,Chen:2023naw}, or complexity \cite{Susskind:2014jwa,Susskind:2014rva,Brown:2015bva} and machine learning \cite{Ahn:2024gjf,Ahn:2024lkh}. An alternative method leverages measurements of pole-skipping points to transform the reconstruction problem from a differential formulation to a system of linear equations, thus facilitating the bulk reconstruction \cite{Lu:2025jgk,Lu:2025pal}. There is also an approach employs a canonical construction of bit threads, formulated in terms of differential forms via the Iyer-Wald formalism, to achieve a local reconstruction of the bulk metric \cite{Agon:2020mvu}.
These methods hypothesize that certain boundary quantities correspond to specific bulk geometric quantities, thus offering ways to recover the bulk metric. Recent advances in data-driven reconstruction \cite{Ahn:2024gjf} and operator algebra approaches \cite{Chen:2023naw} suggest promising pathways in different viewpoints. The foundations of bulk reconstruction trace back to Hamilton, Kabat, Lifschytz, and Lowe's operator reconstruction framework \cite{Hamilton:2006fh}, with subsequent approaches generally following two paradigms. The first assumes a bulk model and iteratively refines it by comparing direct-problem solutions with experimental observables. The second employs mathematical inversion techniques to derive bulk properties directly from boundary measurements. These two paradigms form a logical structure encompassing the various methods discussed earlier. By refining both conceptual understanding and computational strategies, these efforts continue to advance our grasp of how spacetime geometry and gravitational dynamics emerge from quantum field theoretic data in the context of holography.

In fact, the two-point correlation function, as a measurable quantity, has always been a focal point of attention. In the direct process, calculating the two-point correlation function is essentially a scattering problem in the bulk. Thus, it naturally lends itself to the study of holographic reconstruction through the inverse problem of scattering which has been widely studied in quantum mechanics, see \cite{Chadan1989,Kravchenko2020,Kravchenko2021}. In our previous work \cite{Fan:2023bsz}, we demonstrated that the bulk metric of a planar/spherically/hyperbolically symmetric asymptotically anti-de Sitter static black brane/hole can be reconstructed from its boundary frequency two-point correlation functions of two probe scalar operators by solving the Gel'land-Levitan-Marchenko integral equation. Here we noted that the correlation functions of the conserved currents dual to gauge fields differ from the system's optical conductivity by just a factor, so we can obtain the two-point correlation functions corresponding to the gauge fields through measurements and use them to reconstruct the bulk geometry. On the other hand, in the previous work, we required two fields in order to reconstruct the bulk, but in reality, this requirement can be further relaxed.

The problem discussed in this paper focuses on using the inverse problem of scattering to reconstruct the bulk metric and explores its related applications. In this work, we will use scalar fields and gauge fields as concrete examples to illustrate their respective bulk reconstructions and corresponding application scenarios. Our approach begins by analyzing how boundary two-point functions encode bulk scattering data for each field type. For scalar fields, the wave equation in the bulk reduces to a Schr\"{o}dinger-like form after appropriate field redefinitions and coordinate transformations. This allows the use of well-established quantum inverse scattering techniques, such as the Gel'fand-Levitan-Marchenko formalism, to extract effective potentials and ultimately reconstruct the bulk metric function. Notably, for scalar field, we will generalize our previous method and show that correlation function of a single operator is enough to reconstruct the bulk metric of static AdS black holes/branes with planar, spherical, or hyperbolic symmetries. In addition, if we have data of correlation function for two different scalar operators, we can then reconstruct the metric of $(d+1)$-dimensional static homogeneous anisotropic bulk metric. For gauge fields, gauge-invariant combinations of Maxwell equations provide a similar Schr\"{o}dinger structure, albeit with more intricate potential terms due to polarization dependence and gauge constraints. We will show that the data of conductivity could be used to reconstruct the bulk metric in some special cases.

The organization of this paper is as follows: section 2 revisits bulk reconstruction with scalar fields, detailing how momentum dependence resolves limitations of prior methods of \cite{Fan:2023bsz}, and introduces how to reconstruct the geometry with only a single operator. We also explain how to reconstruct the metric for anisotropic metric. Section 3 extends the framework to gauge fields, demonstrating reconstruction for Banados-Teitelboim-Zanelli (BTZ), Schwarzschild-AdS black holes and Reissner-Nordstr\"{o}m (RN)-AdS black hole. Section 4 evaluates the stability of the method under noisy experimental conditions and introduces filtering strategies to enhance accuracy. And finally, we will present a discussion and outlook in Section 5.

\section{Bulk reconstruction with scalar field}

In this section, we will firstly comprehensively review the ``direct problem'' for holographic correlation function of massive scalars, and we will demonstrate how to calculate the two-point correlation functions of the operators that are dual to these fields through the framework of holography. In the process, we will explore the actual relationship between these two-point correlation functions and the bulk geometry. We will obtain an understanding of how these two-point  functions encapsulate information about the bulk geometry. This paper primarily focuses on the linear response theory, under which the back-reaction of the perturbation field on gravity is a higher-order infinitesimal quantity and can be neglected.

\subsection{Reconstruction using the scalars with different momentums}
To better demonstrate how our method obtains the bulk geometry from the correlation function, we first briefly review how to obtain the correlation function on the boundary from the bulk by using two scalar operators. For more detailed steps and processes, please refer to our previous work \cite{Fan:2023bsz}.
\subsubsection*{Reconstruction with two operators}
Consider the free scalar field $\phi$ in the bulk, of which the action reads
\begin{equation}
	S=-\frac{1}{2}\int \td^{d+1}x \sqrt{-g}\left[(\partial\phi)^2+m^2\phi^2\right].
\end{equation}
We consider a stable planar symmetric static asymptotically AdS spacetime. The metric is given as follows:
\begin{equation}\label{btz}
	\td s^2=\frac{1}{z^2(\rho)}\left[h(\rho)(-\td t^2+\td \rho^2)+\sum_{i=d-1}\td x^2_i\right],
\end{equation}
with the following boundary conditions
\begin{equation}
	z(0)=0,\ z'(0)=h(0)=1,\ z(\infty)=z_h,\ h(\infty)=0.
\end{equation}
The horizon is located at $\rho\rightarrow\infty$ and the AdS boundary is set at $\rho=0$. In this work, unless otherwise specified, the inverse horizon $z_h$ is uniformly set as $z_h=1$ in the subsequent calculations. We need to make a perturbation by denoting $\phi=z^{\frac{d-1}{2}}\varphi\te^{-\text{i}\omega t}$ on the fixed background \eqref{btz}, which would give us the equation of motion:
\begin{equation}\label{ode-phi}
	-\partial^2_\rho\varphi+\left(V_\Delta+\frac{4m^2+d^2-1}{4\rho^2}\right)\varphi=\omega^2\varphi,
\end{equation}
where
\begin{equation}\label{poten}
	V_\Delta=-\frac{(d-1)\partial^2_\rho z}{2z}+\frac{d^2-1}{4}\left(\frac{(\partial_\rho z)^2}{z^2}-\frac{1}{\rho^2}\right)+m^2\left(\frac{h(\rho)}{z^2}-\frac{1}{\rho^2}\right).
\end{equation}
We find that Eq.~\eqref{ode-phi} has the same form as the Schr\"{o}dinger equation, and the solution has asymptotical behavior as $\varphi=\varphi^{(s)}\rho^{\Delta_-}(1+\cdots)+\varphi^{(v)}\rho^{\Delta_+}(1+\cdots)$. Here $\Delta_\pm=d/2\pm\sqrt{d^2/4+m^2}$. There are two schemas for quantization. One is called standard quantization which treats $\varphi^{(s)}$ as the source term of boundary theory and $\varphi^{(v)}$ as the value of expectation $\braket{\mathcal{O}}$ for boundary scalar operator $\mathcal{O}$ with conformal dimension $\Delta=\Delta_+$. The other is called alternative quantization which treats $\varphi^{(v)}$ as the source term of boundary theory and $\varphi^{(s)}$ as the value of expectation $\braket{\mathcal{O}}$ for boundary scalar operator $\mathcal{O}$  with conformal dimension $\Delta=\Delta_+$. In the case that $\Delta_->d/2-1$ both schemas of quantization can be applied. However, if $\Delta_-<d/2-1$, then only standard quantization can be applied. We can obtain the two-point correlation function under these different quantizations according to the holographic dictionary. One has
\begin{equation}
	G=
\left\{
\begin{aligned}
	-2\nu\frac{\phi^{(s)}}{\phi^{(v)}} &&\text{alternative quantization},\\2\nu\frac{\phi^{(v)}}{\phi^{(s)}} &&\text{standard quantization}.
\end{aligned}
\right.\label{eq2.6}
\end{equation}
where $\nu=\sqrt{\frac{d^2}{4}+m^2}$. In the direct problem, our task is to find the two-point function $G$ from the given metric~\eqref{btz}.

Now let us consider the inverse problem, i.e., recovering the metric function $\{z(\rho),h(\rho)\}$ from the two-point correlation function $G$. In practice, the two-point correlation function is measured according to the linear response of the system to external perturbations. Then the system should be stable against the small perturbations; otherwise the two-point correlation function cannot be obtained experimentally. Thus, we need to assume that the poles of $G$ all locate at lower complex plane of $\omega$. In this situation, the method from our previous work~\cite{Fan:2023bsz} is applicable. The key aspect of that method lies in reconstructing the potential \eqref{poten}, which can be solved by the following equation; see Ref.~\cite{Chadan1989} for more details.
\begin{equation}\label{effVK1}
  V_{\Delta}(\rho)=2\frac{\td}{\td \rho}K_{\Delta}(\rho,\rho)\,.
\end{equation}
The kernel $K_{\Delta}(\rho, \sigma)$ is determined by the following integral equation:
\begin{equation}\label{GLMeqs}
  K_{\Delta}(\rho,\sigma)+\Omega_{\Delta}(\rho,\sigma)+\int_0^\rho K_{\Delta}(\rho,y)\Omega_{\Delta}(y,\sigma)\td y=0
\end{equation}
and function $\Omega_{\Delta}(\rho,\sigma)$ is given by function $F_{\Delta}(\omega)$ according to
\begin{equation}\label{defOmegav}
  \Omega_{\Delta}(\rho,\sigma)=\int_0^{\infty}\hat{J}_{v}(\omega\rho)\hat{J}_{v}(\omega\sigma)\left(\frac1{|F_{\Delta}(\omega)|^2}-1\right)\td\omega\,.
\end{equation}
Here $\hat{J}_v(x)=\sqrt{x}J_v(x)$ and $J_{v}(x)$ is the $v$-th order the first king of Bessel function. The function $F_{\Delta}(\omega)$ depends on the correlation function $G(\omega)$ according to
\begin{equation}\label{relF2C2}
  |F_{\Delta}(\omega)|^2=\frac{\pi|\omega/2|^{2v}v}{\Gamma(v+1)^2\Im[G_{\Delta}(\omega)]}\,.
\end{equation}

\begin{figure}
   \centering
\begin{tikzpicture}[align=center]
  \node (a) {$F_{\Delta_1}(\omega)$ and $F_{\Delta_2}(\omega)$};
  \node[below=of a] (d) {$\Omega_{\Delta_1}(\rho,\rho)$ and $\Omega_{\Delta_2}(\rho,\rho)$};
  \node[below=of d] (f) {$K_{\Delta_1}(\rho,\rho)$ and $K_{\Delta_2}(\rho,\rho)$};
  \node[below=of f] (g) {$V(\Delta_1,\rho)$ and $V(\Delta_2,\rho)$};
  \node[right=of a] (b) {correlation function \\of the second operator \\$G_{\Delta_2}(\omega)$};
  \node[below=of g] (c) {$z(\rho)$ and $h(\rho)$};
  \node[left=of a] (e) {correlation function \\of the first operator \\$G_{\Delta_1}(\omega)$};

  \draw[->] (g) --  node[right]{\eqref{27} and \eqref{28}}(c);
  \draw[->] (a) --  node[right]{\eqref{defOmegav}}(d);
  \draw[->] (d) --  node[right]{\eqref{GLMeqs}}(f);
  \draw[->] (f) --  node[right]{\eqref{effVK1}}(g);
  \draw[->] (b) --node[above]{\eqref{relF2C2}} (a);
  \draw[->] (e) --node[above]{\eqref{relF2C2}} (a);

\end{tikzpicture}
      \caption{The steps about how to reconstruct the geometry with two operator of two different conformal dimensions. }
      \label{process1}
\end{figure}
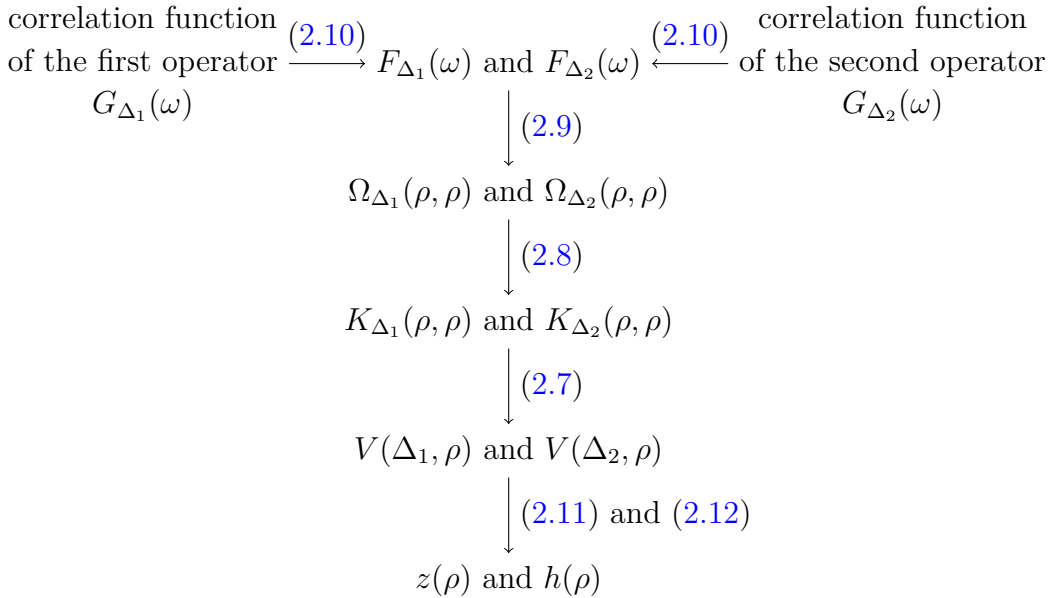

In our previous work \cite{Fan:2023bsz}, we choose two correlation functions with different conformal dimensions $\Delta_1$ and $\Delta_2$, then we can obtain two potentials $V(\Delta_1)$ and $V(\Delta_2)$. From \eqref{poten}, we have:
\begin{equation}\label{27}
	\frac{m^2_2 V(\Delta_1)-m_1^2 V(\Delta_2)}{m_2^2-m_1^2}=\frac{\partial^2_\rho z}{2z}+\frac{1}{4}\frac{(\partial_\rho z)^2}{z^2}.
\end{equation}
By solving the above differential equation, we can recover $z(\rho)$, then we can recover $h(\rho)$ by
\begin{equation}\label{28}
	h=\frac{V(\Delta_1)-V(\Delta_2)}{m_1^2-m_2^2}z^2.
\end{equation}
We can use a flowchart to clarify this process in Fig. \ref{process1}.

\subsubsection*{Reconstruction with single operator}
We have have briefly summarized our previous work~\cite{Fan:2023bsz} on how to reconstruct metric by solving inverse problem of two-point correlation function. It is worth noting that this reconstruction involved two scalar fields (with masses $m_1$ and $m_2$, corresponding to $\Delta_1$ and $\Delta_2$). One hand, measuring linear responses using multiple probes may introduce systematic errors or be affected by interactions between the probes; on the other hand, we might encounter massless probes, in which case no number of probes can yield the two distinct potentials $V(\Delta_1)$ and $V(\Delta_2)$ we aim to obtain.

\begin{figure}
      \centering
      \includegraphics[width=0.49\textwidth]{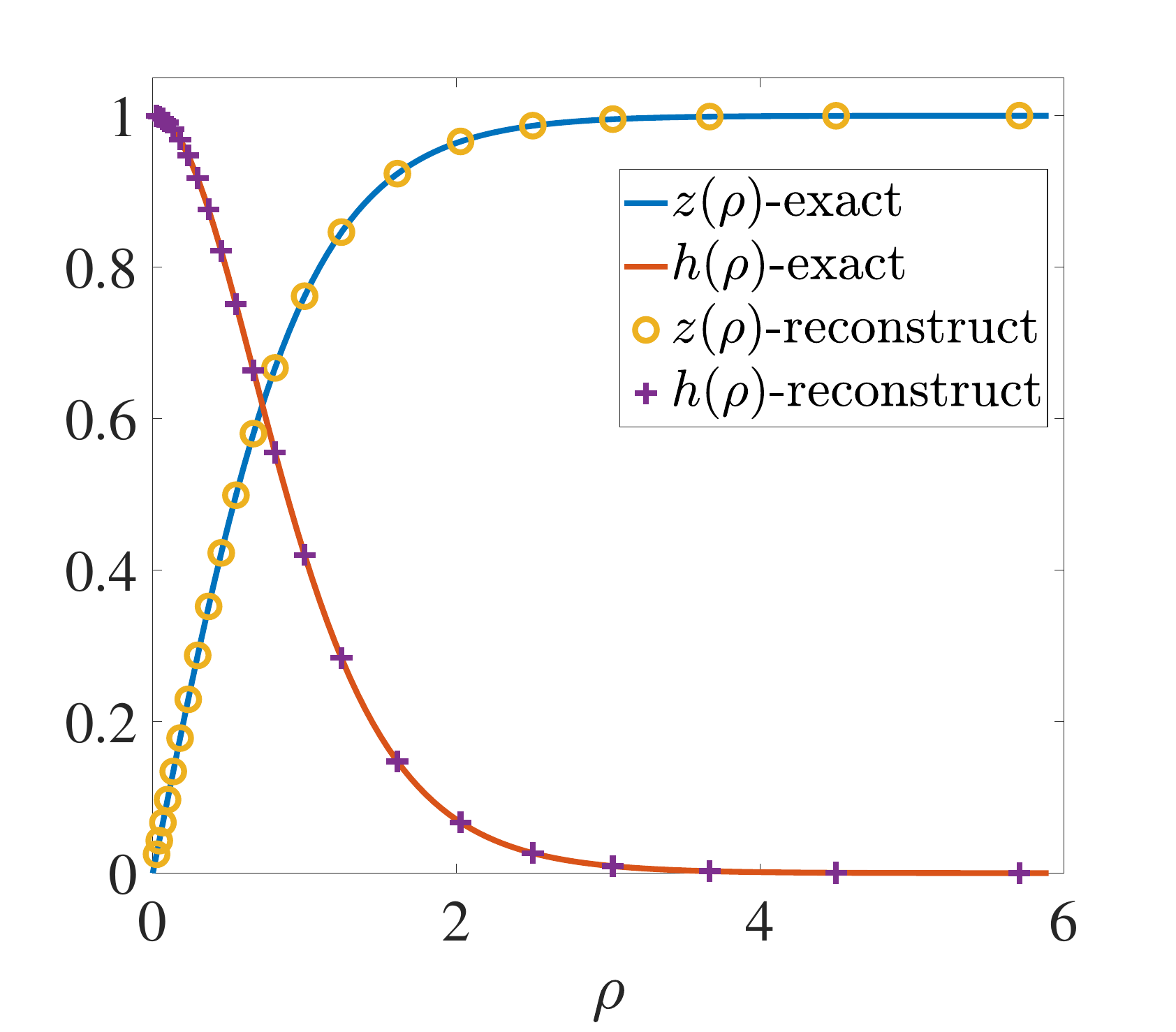}
      \caption{Comparison between the reconstructed metric components and their exact values with our method. Here it uses two-point functions of $k_1=0$ and $k_2=1$ to recover the metric.}
      \label{fig1}
\end{figure}

This raises the question: Is there a way to reduce the number of fields required for the reconstruction? The answer is yes and this will be one main new result in this paper. In fact, we have so far overlooked the dependence of the fields on the momentum $k$. When we account for this dependence on $k$, that is, denoting $\phi=z^{\frac{d-1}{2}}\varphi\te^{-\text{i}\omega t-\text{i}\vec{k}\cdot\vec{x}}$. Now the equation of motion \eqref{ode-phi} remains unchanged, but the potential $V$ \eqref{poten} has been changed into
\begin{equation}\label{potenk}
V_k=-\frac{(d-1)\partial^2_\rho z}{2z}+\frac{d^2-1}{4}\left(\frac{(\partial_\rho z)^2}{z^2}-\frac{1}{\rho^2}\right)+m^2\left(\frac{h(\rho)}{z^2}-\frac{1}{\rho^2}\right)+h(\rho)k^2.
\end{equation}
We can observe an additional term $h(\rho)k^2$, which arises as a consequence of opening the dependence on $k$. This term reflects the impact of considering the momentum dependence, revealing new insights into the reconstruction process, which inspires us that: we can also solve $h(\rho)$ at fixed conformal dimension $\Delta$ and two different momentum $k_1$ and $k_2$ with the following equation:
\begin{equation}\label{h2}
	h=\frac{V_k(k_1)-V_k(k_2)}{k_1^2-k_2^2}.
\end{equation}
Now we take $h(\rho)$ back into \eqref{poten}, and take any one of $V_k(k_1)$ or $V_k(k_2)$ back, thus we can get $z(\rho)$.

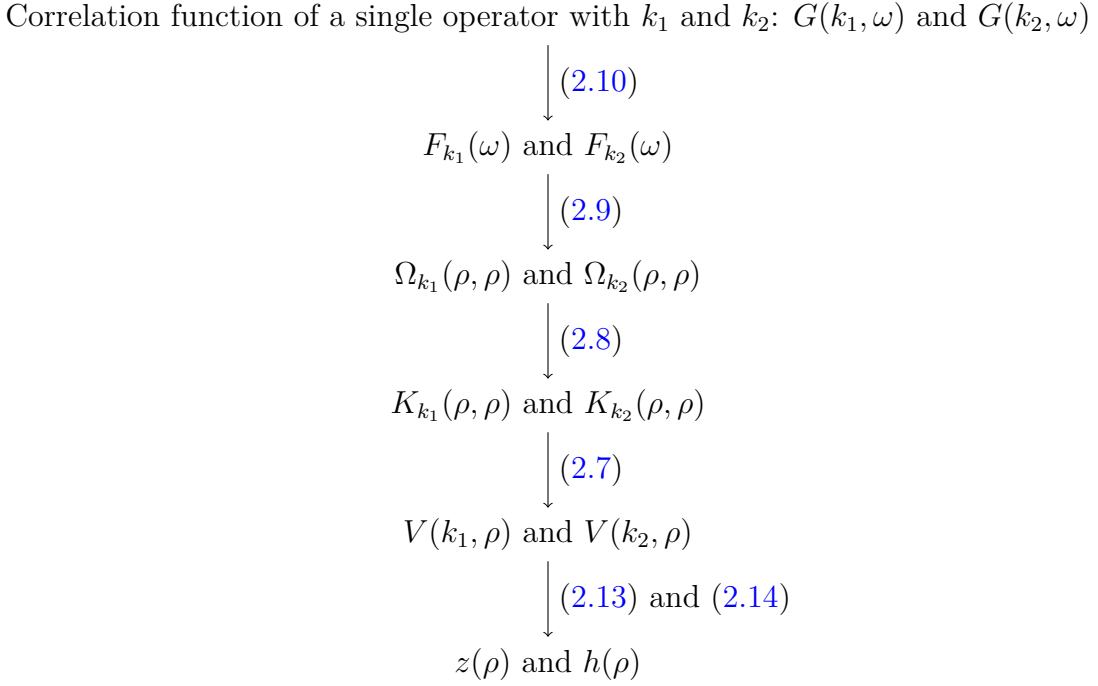
\begin{figure}
   \centering
\begin{tikzpicture}[align=center]
\node (a) {Correlation function of a single operator with $k_1$ and $k_2$: $G(k_1,\omega)$ and $G(k_2,\omega)$};
 \node[below=of a] (b) {$F_{k_1}(\omega)$ and $F_{k_2}(\omega)$};
  \node[below=of b] (d) {$\Omega_{k_1}(\rho,\rho)$ and $\Omega_{k_2}(\rho,\rho)$};
  \node[below=of d] (f) {$K_{k_1}(\rho,\rho)$ and $K_{k_2}(\rho,\rho)$};
  \node[below=of f] (g) {$V(k_1,\rho)$ and $V(k_2,\rho)$};
  \node[below=of g] (c) {$z(\rho)$ and $h(\rho)$};

  \draw[->] (g) --  node[right]{\eqref{potenk} and \eqref{h2}}(c);
  \draw[->] (b) --  node[right]{\eqref{defOmegav}}(d);
  \draw[->] (d) --  node[right]{\eqref{GLMeqs}}(f);
  \draw[->] (f) --  node[right]{\eqref{effVK1}}(g);
  \draw[->] (a) --node[right]{\eqref{relF2C2}} (b);

\end{tikzpicture}
      \caption{The steps on how to reconstruct the geometry with a single operator. }
      \label{process2}
\end{figure}

Let us now consider an example: the reconstruction of the BTZ black hole metric using different values of $k$. For the BTZ black hole, there exists: $z(\rho)=\tanh{\rho}$, and $h(\rho)=1/\cosh^2{\rho}$. At this point, the two-point correlation function of the scalar operator takes the following form:
\begin{equation}\label{2pfkw}
    G(k,\omega)=\frac{\Gamma(-\nu)\Gamma(\frac{1+\nu-\text{i}\omega+\text{i}k}{2})\Gamma(\frac{1+\nu-\text{i}\omega-\text{i}k}{2})}{\Gamma(\nu)\Gamma(\frac{1-\nu-\text{i}\omega+\text{i}k}{2})\Gamma(\frac{1-\nu-\text{i}\omega-\text{i}k}{2})}.
\end{equation}
The poles of 2-point correlation function are setting at $\omega_l=-\text{i}(1+\nu+2l)\pm k$ with $l\in\mathbb{N}^+$ and $k\in\mathbb{R}$. Thus we can say that the poles of $G$ all locate at the lower complex of $\omega$. From this, we can obtain the reconstructed metric of the BTZ black hole. Here, we used the two-point correlation functions for $k = 0$ and $k = 1$, respectively. The comparison between the exact BTZ metric and its reconstruction from the correlation function is shown in Fig.~\ref{fig1}. We also use a flowchart to clarify this process in Fig. \ref{process2}. It is worth noting that although the dependence on $k$ is introduced here, the form of equations \eqref{effVK1}, \eqref{GLMeqs}, \eqref{defOmegav} and \eqref{relF2C2} remains unchanged; it merely includes a dependence on $k$. Therefore, we will not rewrite them here.

This method, which leverages the dependence on $k$ to use two $k$ values instead of two $\Delta$, is particularly useful for massless fields. For the massless field, we can only use the second method to obtain $h(\rho)$ from \eqref{h2}, and then take it back into one's potential to get $z(\rho)$.

\subsection{Bulk reconstruction for an anisotropic static metric}
In the previous examples, we focused on planar symmetric, static, asymptotically AdS spacetimes. However, our method can, in fact, be applied with less restrictive symmetry requirements. Theoretically, it is valid for any static asymptotically AdS spacetime. In such a spacetime the off-diagonal components of the metric are all zero, and the metric depends only on the radial coordinate $z$, not on the time coordinate $t$ or other spatial coordinates $x^i$. In this section, we focus on the following general anisotropic static metric:
\begin{equation}\label{general}
	\td s^2=-g_{tt}(z)\td t^2+g_{zz}(z)\td z^2+\sum_{i}g_{ii}(z)\td x^i\td x^i\,.
\end{equation}
\subsubsection{Method to reconstruct metric}
Now we introduce how to obtain the 2-point correlation function of a scalar under the metric \eqref{general}. The equation of motion for $\phi$ can be written in the following form:
\begin{equation}
	\left( g^{zz}\partial^2_z\phi+\frac{\partial_z \sqrt{-g} g^{zz}\partial_z\phi}{\sqrt{-g}}\right)-g^{tt}\partial^2_t\phi+\sum_{i}g^{ii}\partial_i^2\phi-m^2\phi=0,
\end{equation}
where $i$ represents for coordinates except for $t$ and $z$. After a Fourier transformation by $t\rightarrow\omega$ and $x_i\rightarrow k_i$ we obtain
\begin{equation}\label{eqforgzi}
    \partial^2_z\phi+\frac{\partial_z \sqrt{-g}g^{zz}}{\sqrt{-g}g^{zz}}\partial_z\phi-\left(\sum_{i}\frac{g^{ii}k_i^2}{g^{zz}}-\frac{g^{tt}\omega^2}{g^{zz}}+\frac{m^2}{g^{zz}}\right)\phi=0.
\end{equation}
In order to translate this equation into Schr\"{o}dinger's form, we denote
\begin{equation}\label{s2z}
    s=\int_0^z\sqrt{\frac{g^{tt}}{g^{zz}}}\td x.
\end{equation}
The Eq.~\eqref{eqforgzi} then becomes
\begin{equation}\label{eqforgzi2}
    \partial^2_s\phi+\left(\frac{\partial_z \sqrt{-g}g^{zz}}{\sqrt{-g}g^{zz}}\cdot\sqrt{\frac{g^{zz}}{g^{tt}}}-\frac{\partial}{\partial z}\frac{g^{zz}}{g^{tt}}\right)\partial_s\phi-\left(\sum_{i}\frac{g^{ii}k_i^2}{g^{tt}}-\omega^2+\frac{m^2}{g^{tt}}\right)\phi=0.
\end{equation}
We now see that the $\omega^2$ term no longer contains does not mix with any other function and so can be regarded as an eigenvalue. In order to further deform above equation into  Schr\"{o}dinger's form, we make another transformation $\varphi(s)\psi(s)=\phi(s)$, where $\varphi$ is given by
\begin{equation}\label{eqforvarphi}
	2\partial_s\varphi+\left(\frac{\partial_z\sqrt{-g} g^{zz}}{\sqrt{-g}g^{zz}}\cdot\sqrt{\frac{g^{zz}}{g^{tt}}}-\frac{\partial}{\partial z}\frac{g^{zz}}{g^{tt}}\right)\varphi=0.
\end{equation}
Then we can obtain
\begin{equation}\label{eqforgzi3}
    -\partial_s^2\psi+V_{\text{eff}}\psi=\omega^2\psi,
\end{equation}
where $V_{\text{eff}}$ is given by
\begin{equation}\label{veff}
    V_{\text{eff}}=\sum_{i}\frac{g^{ii}k_i^2}{g^{tt}}+\frac{m^2}{g^{tt}}-\frac{\partial_s^2\varphi}{\varphi}+2\left(\frac{\partial_s\varphi}{\varphi}\right)^2.
\end{equation}
The asymptotic solution of Eq.~\eqref{eqforgzi3} is as follows:
\begin{equation}
    \psi=\psi_1 s^{d-\Delta}(1+\cdots)+\psi_2 s^\Delta(1+\cdots),
\end{equation}
here $\Delta=d/2+\sqrt{(d/2)^2+m^2}$. And from the holographic dictionary, the 2-point correlation function is given by $G\sim{\psi_2}/{\psi_1}$.
The basic steps of reconstructing metric then are as follows:
\begin{enumerate}
\item[Step 1] We fix values of all momentums $k_i$ and use the correlation functions of two different conformal dimensions $\Delta_1$ and $\Delta_2$ to reconstruct the effective potentials $V_{\text{eff}}(\Delta_1)$, $V_{\text{eff}}(\Delta_2)$;
\item[Step 2] We use Eq.~\eqref{veff} to solve the metric component $g_{tt}$ from the values of $V_{\text{eff}}(\Delta_1)$, $V_{\text{eff}}(\Delta_2)$:
\begin{equation}\label{gttgiis1}
	g_{tt}=\frac{V_{\text{eff}}(\Delta_1)-V_{\text{eff}}(\Delta_2)}{m_1^2-m_2^2}.
\end{equation}
\item[Step 3] For every spatial coordinate index $i$, we use the correlation functions of two different momentums $k_i^{(1)}$ and $k_i^{(2)}$ with a same conformal dimension to reconstruct the effective potentials $V_{\text{eff}}(k_i^{(1)})$, $V_{\text{eff}}(k_i^{(2)})$;
\item[Step 4] We use Eq.~\eqref{veff} to solve the metric component $g_{ii}$ from the values of $V_{\text{eff}}(k_i^{(1)})$, $V_{\text{eff}}(k_i^{(2)})$ as follows:
\begin{equation}\label{gttgiis2}
	g_{ii}=g_{tt}\frac{{k_i^{(1)}}^2-{k_i^{(2)}}^2}{V_{\text{eff}}(k_i^{(1)})-V_{\text{eff}}(k_i^{(2)})}.
\end{equation}
Moreover, it is worth noting that when we calculate a specific $k_i $, such as $i=1 $, we not only need to fix $\Delta $, but also fix all $k_ {j \ne 1} $.
\item[Step 5] We then use $g_{tt}$, all $g_{ii}$, any one $V_{\text{eff}}$ that we have obtained in above steps to find function $\varphi$ according to Eq.~\eqref{veff};
\item[Step 6] Finally, we use the function $g_{tt}$ and $\varphi$ to obtain metric component $g_{zz}$ by solving Eq.~\eqref{eqforvarphi};
\end{enumerate}

The metric components obtained here are functions of the new coordinate $s$ rather than $z$. However, we can use the inverse of the coordinate transformation~\eqref{s2z} to recover the dependence on $z$. It is important to note that here we only used $G(\vec{k},\omega)$ of two scalar fields with different masses.

\subsubsection{An example}
\begin{figure}
      \centering
      \includegraphics[width=0.49\textwidth]{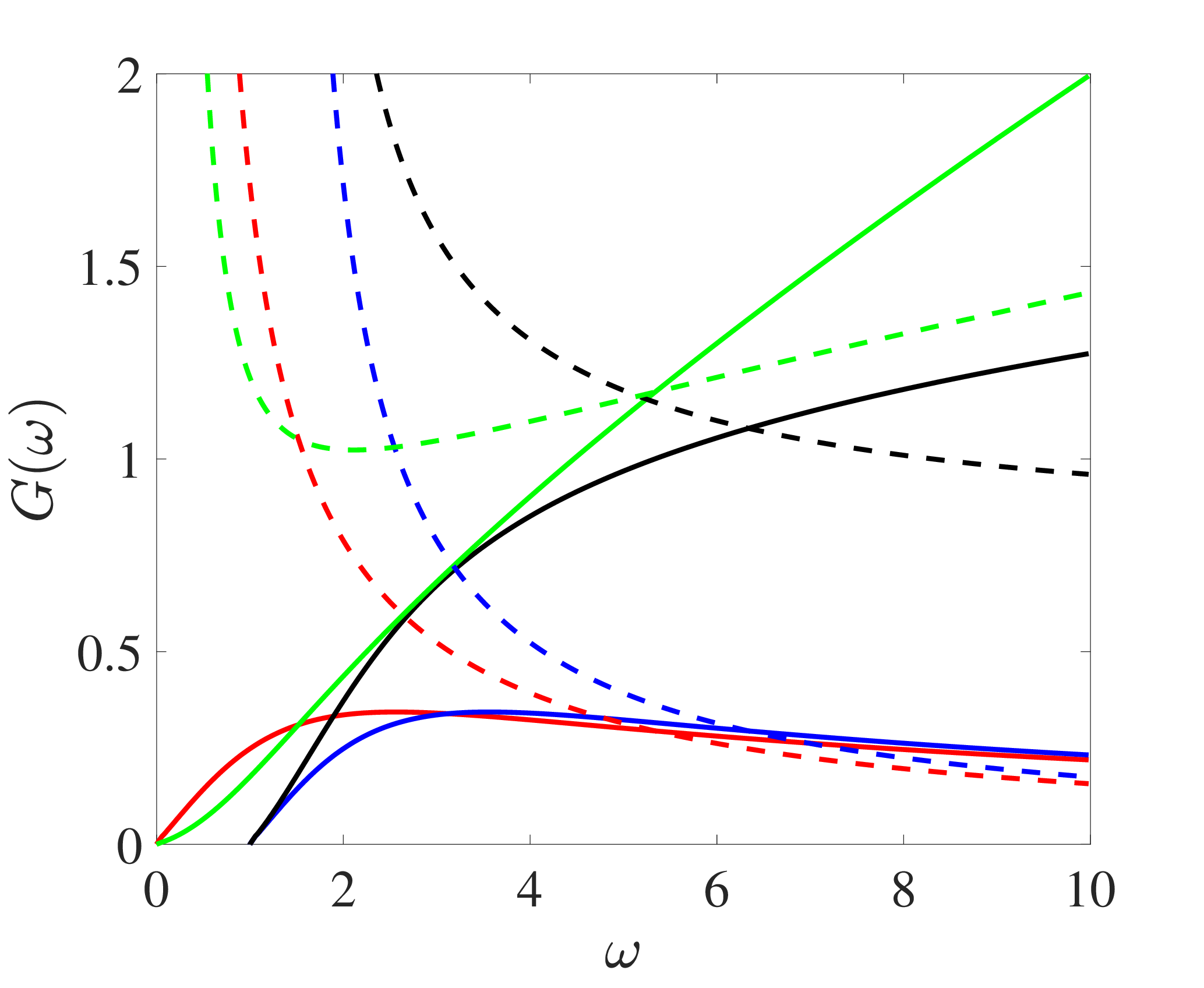}
      \includegraphics[width=0.49\textwidth]{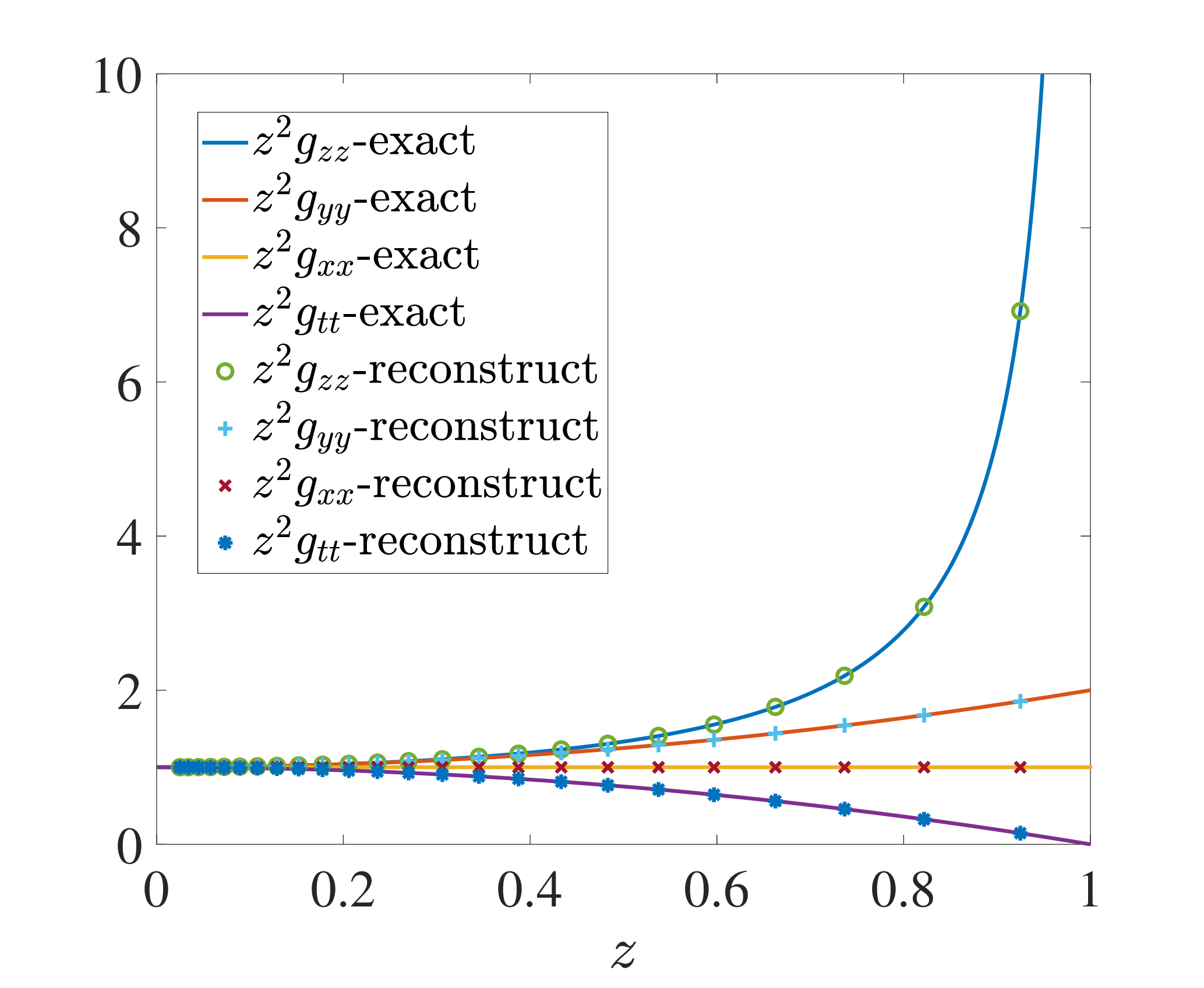}
      \caption{Left panel: correlation functions for $(\Delta,k_x,k_y)=$(4,0,0)(red), (4,1,0)(blue), (4,0,1)(black) and (3,0,0)(green). The solid lines correspond to the imaginary part of the correlation functions and the dashed lines correspond to the real part of the correlation functions. Right panel: the reconstructed components of metric and the comparison to their exact results.}
      \label{gmn}
\end{figure}

As an example, we can write a 4-dimensional anisotropic black hole:
\begin{equation}\label{gans}
    g_{zz}=\frac{1}{z^2(1-z^2)},\ g_{tt}=\frac{1-z^2}{z^2},\ g_{xx}=\frac{1}{z^2},\ g_{yy}=\frac{1+z^2}{z^2}.
\end{equation}
We here do not state that this metric describes spacetime that is dual to any real strongly coupled boundary field in holography. It is only a pedagogical example to exhibit how our above method would successfully reconstruct metric from the data of correlation functions. The metric is just one simple example for our this purpose.

Before showing how our above method are used in reconstruct metric, we first use such a metric to calculate the two-point correlation functions and the results are shown in the left panel of Fig.~\ref{gmn}. Here we choose for $(\Delta,k_x,k_y)=(4,0,0), (4,1,0)$, $(4,0,1)$ and $(3,0,0)$. By making use of these correlation functions, we can then reconstruct the metric. In the step 1, we use the correlation function of $k_x = k_y = 0$ and $\Delta = 4, 3$, reconstruct $V_{\text{eff}}(\Delta=4,k_x=k_y=0)$ and $V_{\text{eff}}(\Delta=3,k_x=k_y=0)$. We then use Eq.~\eqref{gttgiis1} in the step 2 to obtain $g_{tt}$. In the step 3, we use the correlation function of $\Delta=4$ and $k_x = 1, k_y=0$ to reconstruct $V_{\text{eff}}(\Delta=4,\ k_y=0,k_x=1)$. Then we follow the step 4 to  obtain $g_{xx}$ according to  Eq.~\eqref{gttgiis2}. Similarity, we can use the correlation functions of $(\Delta=4,\ k_x=k_y=0)$ and $(\Delta=4,k_x=0,\ k_y=1)$ to reconstruct $V_{\text{eff}}(\Delta=4,\ k_x=k_y=0)$ and $V_{\text{eff}}(\Delta=4,k_x=0,\ k_y=1)$ and so obtain obtain  $g_{yy}$ as well. We can following the step 5 to find the function $\varphi$ and follow the step 6 to obtain the metric component $g_{zz}$. Note all above metric components are reconstructed from the coordinate $s$. We transform them into coordinate $z$ according to Eq.~\eqref{s2z}. Fig.~\ref{gmn} illustrates the dependence of the three reconstructed metric components $g_{zz}$, $g_{tt}$, and $g_{xx}$, $g_{yy}$ on the $z$ coordinate, along with a comparison to the exact results. Although only a simple example is presented, we can still observe that our method expands the range of models it can address.



\section{Bulk reconstruction with gauge field}

In the previous section, we introduced the method of bulk reconstruction using the inverse scattering problem with scalars . However, as we mentioned, our method is not only useful for scalars; in fact, it remains applicable to gauge fields as well. In this section, we will begin discussing the bulk reconstruction problem for gauge fields and its applications. Gauge fields are fundamental to understanding various physical phenomena, particularly in the context of charged black holes and their interactions with matter fields in asymptotically AdS spacetimes. Before delving into the details of bulk reconstruction for gauge fields, it is useful to review the gravity theory's action that governs the dynamics of gauge field in this framework. The action of a gauge field in asymptotically AdS spacetime is given by
\begin{equation}\label{vecact}
	S=\int \td^{d+1}x \sqrt{-g}\left(-\frac{1}{4}F_{\mu\nu}F^{\mu\nu}\right),
\end{equation}
here $F_{\mu\nu}=(\td A)_{\mu\nu}$ is the field strength tensor of the Maxwell field, and  $A_\mu$ is the U(1) gauge field. According to the holographic dictionary, global conserved current $J$ on the boundary corresponds to gauge field $A_\mu$ in the bulk. Therefore, we can study the correlation functions of conserved currents on the boundary through the gauge fields. Correspondingly, we can conversely utilize the measurable two-point correlation functions of conserved currents to reconstruct the bulk theory.

There is a correspondence between the two-point correlation function
$G$ of the conserved current $J$ and the optical conductivity $\sigma$ of the material, as shown by the following relationship:
\begin{equation}\label{sigma}
    \sigma(\omega,\vec{x})=\frac{\mathcal{G}(\omega,\vec{x})}{\text{i}\omega}.
\end{equation}
Here $\omega$ is the frequency and $\mathcal{G}(\omega,\vec{x})$ is related to the momentum correlation function $G(\omega,\vec{k})$ by the inverse Fourier transformation $\mathcal{G}(\omega,\vec{x})=(2\pi)^{-d/2}\int G(\omega,\vec{k})e^{i\vec{k}\cdot\vec{x}}\td^{d}x$. Optical conductivity is a physically measurable quantity that can be used to reconstruct the bulk geometry by measuring the conductivity of the corresponding material. After one obtains the optical conductivity $\sigma(\omega,\vec{x})$, one can perform Fourier transformation by converting position-dependence into momentum-dependence and obtain correlation function in frequency and momentum space. Since the optical conductivity is an important and common measurable quantity of materials, Eq.~\eqref{sigma} shows that the solving inversion problem for correlation function of gauge field will have more real significance.

While the scalar field provides a foundational framework for bulk reconstruction, gauge fields play a pivotal role in capturing the dynamics of charged systems, such as charged black holes and holographic superconductors. Unlike scalar fields, gauge fields inherently lack a mass term, imposing unique constraints on their reconstruction. Thus, we cannot following the previous of work~\cite{Fan:2023bsz}, which used two fields with different masses to reconstruct bulk metric. This section extends our inverse scattering method to gauge fields, addressing both their direct correlation functions and the challenges posed by their zero mass. We first review the holographic dictionary for gauge fields, emphasizing how their two-point functions encode bulk geometry. Subsequently, we demonstrate that momentum dependence-analogous to the role of conformal dimension in scalars-enables bulk reconstruction even for massless fields. As we demonstrated in the previous section, even with only a single field, activating the momentum dependence allows for the reconstruction of the metric. Through concrete examples, we illustrate the universality of our approach across different symmetries and dimensions.

\subsection{Correlation function of gauge field in natural background}
To discuss the inverse problem for gauge fields, let us review its direct problem. Similarly, the two-point correlation function of the conserved current associated with the gauge field can also be obtained through experimental measurements. It is helpful to simplify the system by focusing on only the dynamics of the gauge field in order to bridge the discussion between the general role of gravity theory and Maxwell's equations.  The Maxwell equations are given by:
\begin{equation}\label{Maxwell}
	\partial_{\mu}\left[\sqrt{-g}(\partial^\mu A^{\nu}-\partial^{\nu} A^{\mu})\right]=0 .
\end{equation}
Since the background spacetime is still assumed to be static planar symmetric back hole, we still use coordinate gauge~\eqref{btz}.  Here we first consider the \emph{natural} black hole, which means that the black hole does not contains charge, thus $A_t=0$. The boundary theory is dual to a natural black hole.

We consider the conductivity of material under small external electric field. Holographically, we then need to study the linear responded theory of probe Maxwell field in a natrual black hole. Since the background is assumed to be neutral with $A_\mu=0$, the linear perturbational theory of gauge field become simple. Only consider the $x$-component of the gauge fields, add a perturbation $A_{x}=z^{(d-3)/2}\mathcal{A}_{x}(\rho)\te^{\text{i}\omega t-\text{i}ky}$ on this system (here we take $x_1=x$ and $x_2=y$), thus \eqref{Maxwell} can become
\begin{equation}\label{mxwzk}
  -\frac{\td^2\mathcal{A}_{x}}{\td\rho^2}+\left(V_{k}+\frac{d^2-4d+3}{4\rho^2}\right)\mathcal{A}_{x}=\omega^2\mathcal{A}_{x}.
\end{equation}
Here
\begin{equation}\label{mxwzkV}
\begin{split}
  V_{k}&=-\frac{(d-3)z''}{2z}+\frac{(d^2-4d+3)}4\left(\frac{z'^2}{z^2}-\frac1{\rho^2}\right)+k^2h.
  \end{split}
\end{equation}
The reason we activate the dependence on $k$ here is that if we do not introduce this dependence, the coefficient of the $h$-term in \eqref{mxwzkV} would vanish. This would imply that the potential no longer depends on $h$, making it impossible to reconstruct the metric function $h$. Therefore, activating the $k$-dependence is necessary.

The solution to these ordinary partial equations are similar to the equation of scalars and we can find near the conformal boundary the solution takes the form
\begin{equation}
	\mathcal{A}_x=\mathcal{A}_{x}^{-} {\rho}^{\Delta_-}(1+\cdots)+\mathcal{A}_{x}^{+} {\rho}^{\Delta_+}(1+\cdots),
\end{equation}
where $\Delta_\pm$ is characteristic exponent of the perturbation equation, and $\Delta_-<\Delta_+$. The dots will vanish as $z(\rho)\rightarrow 0$. In fact, they are the conformal dimensions of the boundary operator $J$ dual to $A$. Different from the scalar field, for the gauge field, we have $(\Delta-1)(\Delta-d+1)=0$. Here $\Delta_-=1$ and $\Delta_+=d-1$. According to the holographic dictionary, we can know $A_x^-$ is the source of the dual operator, and $A_x^+$ is its expectation value. According to the holographic dictionary, the relationship between the two-point correlation function of the conserved current $J$ and the gauge field is given by:
\begin{equation}\label{GJ}
    G(J)=-\frac{\mathcal{A}^+}{\mathcal{A}^-}.
\end{equation}
Combine \eqref{sigma} with \eqref{GJ}, we will obtain the relationship between the optical conductivity $\sigma$ and gauge fields $A$:
\begin{equation}
    \sigma=-\frac{\mathcal{A}^+}{\text{i}\omega \mathcal{A}^-}
\end{equation}
In the following calculations, we will consistently utilize this relationship and briefly review the direct problems of gauge field in BTZ black hole and Schwarzschild-AdS black hole.

\subsubsection*{BTZ black hole}
Firstly, let us focus on the BTZ black hole. In this case we have $d=2$, $\Delta_\pm=1$. This  will lead to a logarithmic term in the $\Delta_-$ branch, i.e. $A_x=A_x^-\ln{\Lambda z}+A_x^++\cdots$. Here $\Lambda$ is a renormalization scale. In fact, we can shift the two-point function with a constant $\ln{\Lambda}$ because of the logarithmic term. We find that the logarithmic term is the leading term. Now try to calculate its boundary frequency two-point function and the frequency-dependent conductivity. Consider the metric \eqref{btz} and set $d=2$, for the BTZ black hole we have $z(\rho)=\tanh(\rho)$ and $h=1/\cosh^2(\rho)$. Near the AdS boundary ($\rho\rightarrow0^+$), the asymptotic behavior of $\mathcal{A}_{x}$ take the form $\mathcal{A}_x=\mathcal{A}_x^-\ln{\Lambda \rho}+\mathcal{A}_x^++\cdots$. Solve \eqref{mxwzk}, we obtain
\begin{equation}
	\begin{split}
		\mathcal{A}_x&=C_1 \sqrt{\sinh{\rho}}(\cosh{\rho})^{\frac{1}{2}-\text{i}\sqrt{k^2+1}}{}_2F_1\left(\tilde{a},\tilde{b},\tilde{a}+\tilde{b},\cosh^2{\rho}\right)\\
		&+C_2 \sqrt{\sinh{\rho}}(\cosh{\rho})^{\frac{1}{2}+\text{i}\sqrt{k^2+1}}{}_2F_1\left(\tilde{a}^*,\tilde{b}^*,\tilde{a}^*+\tilde{b}^*,\cosh^2{\rho}\right),
	\end{split}
\end{equation}
where ${}_2F_1(x)$ is the hypergeometric function, and $\tilde{a}=(1-\text{i}\sqrt{k^2+1}+{\text{i}\omega})/{2}$ and $\tilde{b}=(1-\text{i}\sqrt{k^2+1}-{\text{i}\omega})/{2}$. Impose the incoming-wave boundary condition at the horizon, the frequency and momentum conductivity take the form
\begin{equation}\label{sigkw}
	\sigma(\omega)=\frac{\text{i}}{\omega}\left[\gamma+2\pi\text{i}+\frac{1}{2}\psi\left(\frac{\text{i}\omega-\sqrt{1+k^2}}{2}\right)+\frac{1}{2}\psi\left(\frac{\text{i}\omega+\sqrt{1+k^2}}{2}\right)+\ln{\Lambda}\right].
\end{equation}
where and $\psi(x)=\Gamma'(x)/\Gamma(x)$ is the digamma function. In fact, there is a pole at $\omega=0$ in the imaginary part of the conductivity from the digamma function, and it corresponds to a delta function at $\omega=0$ in the real part, and it was the Kramers-Kronig relation. We find the DC conductivity is infinity, which arises from the system's translational symmetry, and it does not correspond to the holographic superconductivity as noted in \cite{Hartnoll:2008kx,Ren:2010ha}.

\subsubsection*{The $4$-dimensional situation}
In the $4$-dimensional situation, \eqref{mxwzk} takes the form that
\begin{equation}
    -\frac{\td^{2}}{\td\rho^{2}}{\mathcal{A}_x}+k^2 h{\mathcal{A}_x}= \omega^{2}{\mathcal{A}_x}
\end{equation}
It can be noted that the dependence on the function $z(\rho)$ has entirely disappeared from this equation. This is because the four-dimensional Maxwell theory is conformally invariant, and $z(\rho)$ in the metric \eqref{btz} acts as a conformal factor that remains free in the gauge field description. Thus, in the case of a $4$-dimensional bulk theory, reconstruction solely through gauge fields can only recover the function $h(\rho)$. To reconstruct the full bulk geometry, additional fields (such as scalar fields) must be incorporated.
\subsubsection*{Schwarzschild-AdS black hole}
For the 5-dimensional case, we consider the Schwarzschild-AdS black hole. In coordinate system $\{z,t,x,y,u\}$, its metric can be written as:
\begin{equation}
     \td s^2=\frac{1}{z^2}\left[-f(z)\td t^2 + \frac{\td z^2}{f(z)} + \td x^2 + \td y^2 + \td u^2\right],
\end{equation}
where $f(z)$ could be set as
\begin{equation}
    f(z)=1-z^4.
\end{equation}
Here the horizon is set at $z=1$. After applying coordinate transformation $\{z\rightarrow\rho\}$, we obtain a metric of the form \eqref{btz}, where $h(\rho)=1-z^4(\rho)$ and $z(\rho)$ does not have an explicit analytic form, and its correspondence with $\rho$ is determined by an integral equation. After simplification, it reduces to the following algebraic equation:
\begin{equation}
    \rho=\frac{1}{2}\left(\tan^{-1}{z}+\tanh^{-1}{z}\right).
\end{equation}
\begin{figure}
      \centering
      \includegraphics[width=0.55\textwidth]{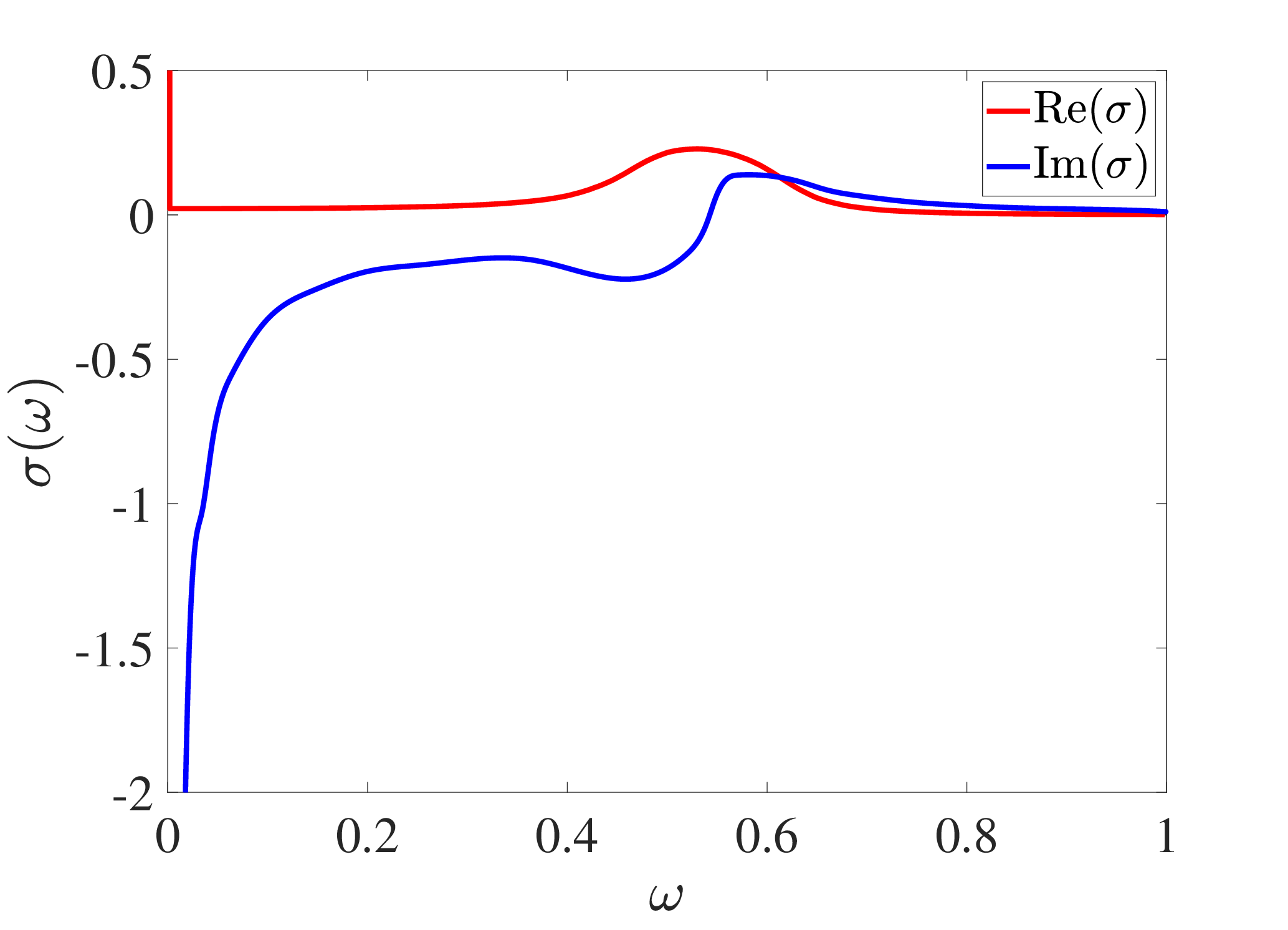}
      \caption{The conductivity of Schwarzschild-AdS black hole, there are real and imaginary parts. Here we choose $k=0.5$. The pole in the imaginary part at $\omega=0$ corresponds to a $\delta$-function in the real part.}
      \label{swz}
\end{figure}

In this case there is no compact expression for the conductivity and we can only obtain it numerically. We solve the equation of motion \eqref{mxwzk} under these conditions and derive the conductivity based on the holographic dictionary, here we set $k=0.5$ and the result is shown in Fig.\ref{swz}. As mentioned earlier, the real and imaginary parts of the conductivity exhibit Kramers-Kronig relationship. The translation invariant also leads to that DC conductivity is infinity.

\subsection{Inverse problem for gauge field in natural background}
In the previous sections, we reviewed how to solve the direct problem for gauge field in natural background: namely, calculating the two-point correlation functions of the conserved current from the equations of motion for the gauge fields. It can be observed that, aside from the addition of a few extra equations and except for $d=3$, the structure of the individual components for gauge fields does not significantly differ from that of scalar fields. Consequently, reconstructing the geometry based on each independent component is similarly straightforward. We now turn to the inverse problem: reconstructing the metric from boundary conductivity data. By analyzing conductivity measurements at distinct momentum values $k_1$ and $k_2$, we disentangle the metric components $h(\rho)$ and $z(\rho)$, mirroring the scalar field reconstruction. More detailed, the solving inverse problem contains following two steps:
\begin{enumerate}
\item[Step 1] We choose two different  momentums $k_1$ and $k_2$ to reconstruct the effective potentials  $V_{k_1}$ and $V_{k_2}$;
\item[Step 2] Solve the metric component $h$ via
\begin{equation}\label{mathcalH}
    h(\rho)=\frac{V_{k_1}-V_{k_2}}{k^2_1-k^2_2}.
\end{equation}
\item[Step 3] Take $h$ and $V_{k_1}$ (or $V_{k_2})$ into Eq.~\eqref{mxwzkV} and we can solve the metric component $z(\rho)$.
\end{enumerate}
This method not only circumvents the limitations of massless fields but also highlights the adaptability of our framework to diverse field types. We validate this approach through reconstructions of the BTZ and Schwarzschild-AdS geometries, comparing numerical results with exact solutions to demonstrate robustness. By performing calculations on these two models, we will demonstrate the effectiveness of our method. As previously discussed, in the four-dimensional case, the conformal invariance of Maxwell theory leads to the inability to reconstruct the conformal factor $z(\rho)$ through gauge fields. Therefore, we will not present the reconstruction for the four-dimensional theory here.
\subsubsection*{BTZ black hole}
\begin{figure}
      \centering
      \includegraphics[width=0.49\textwidth]{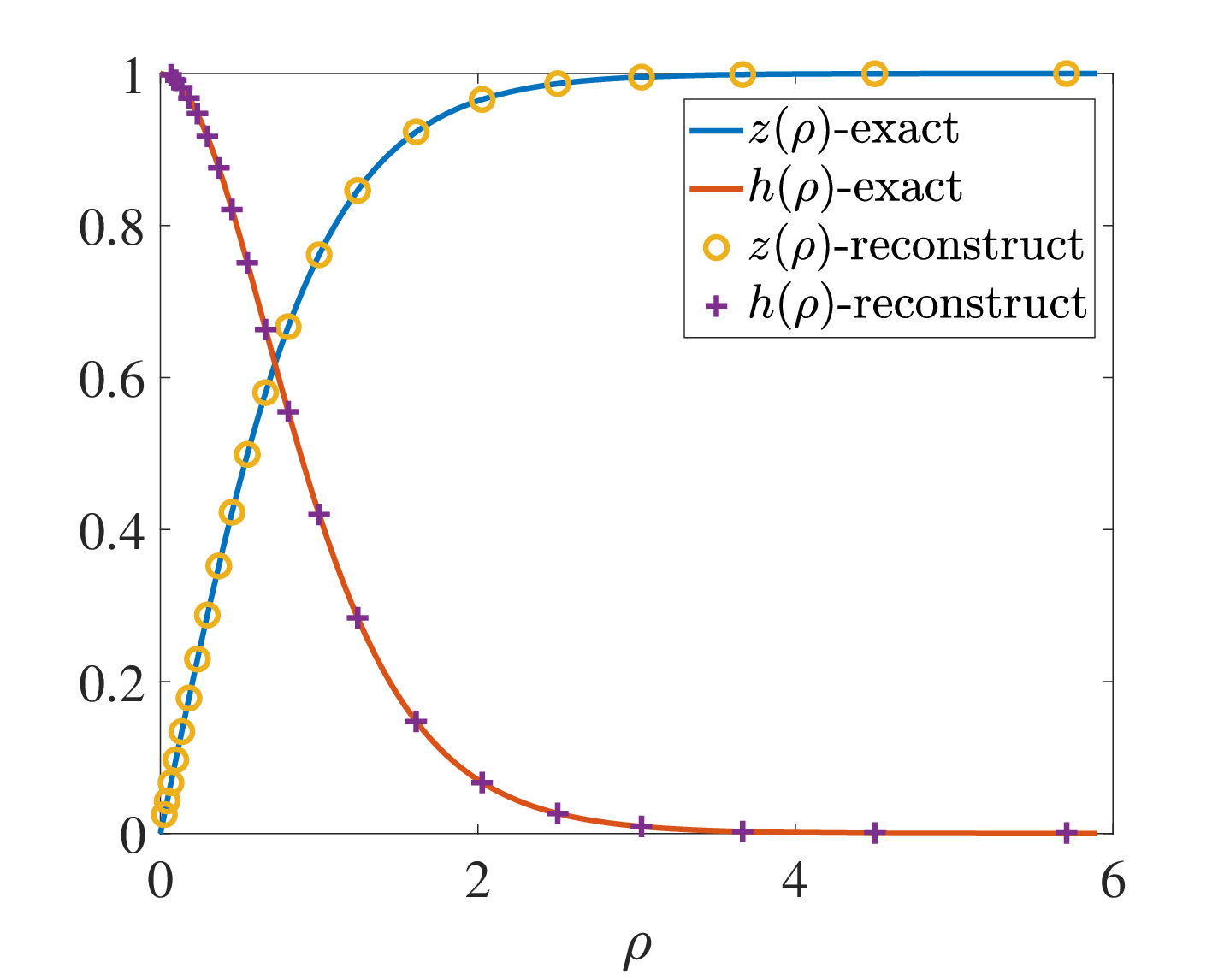}
      \includegraphics[width=0.49\textwidth]{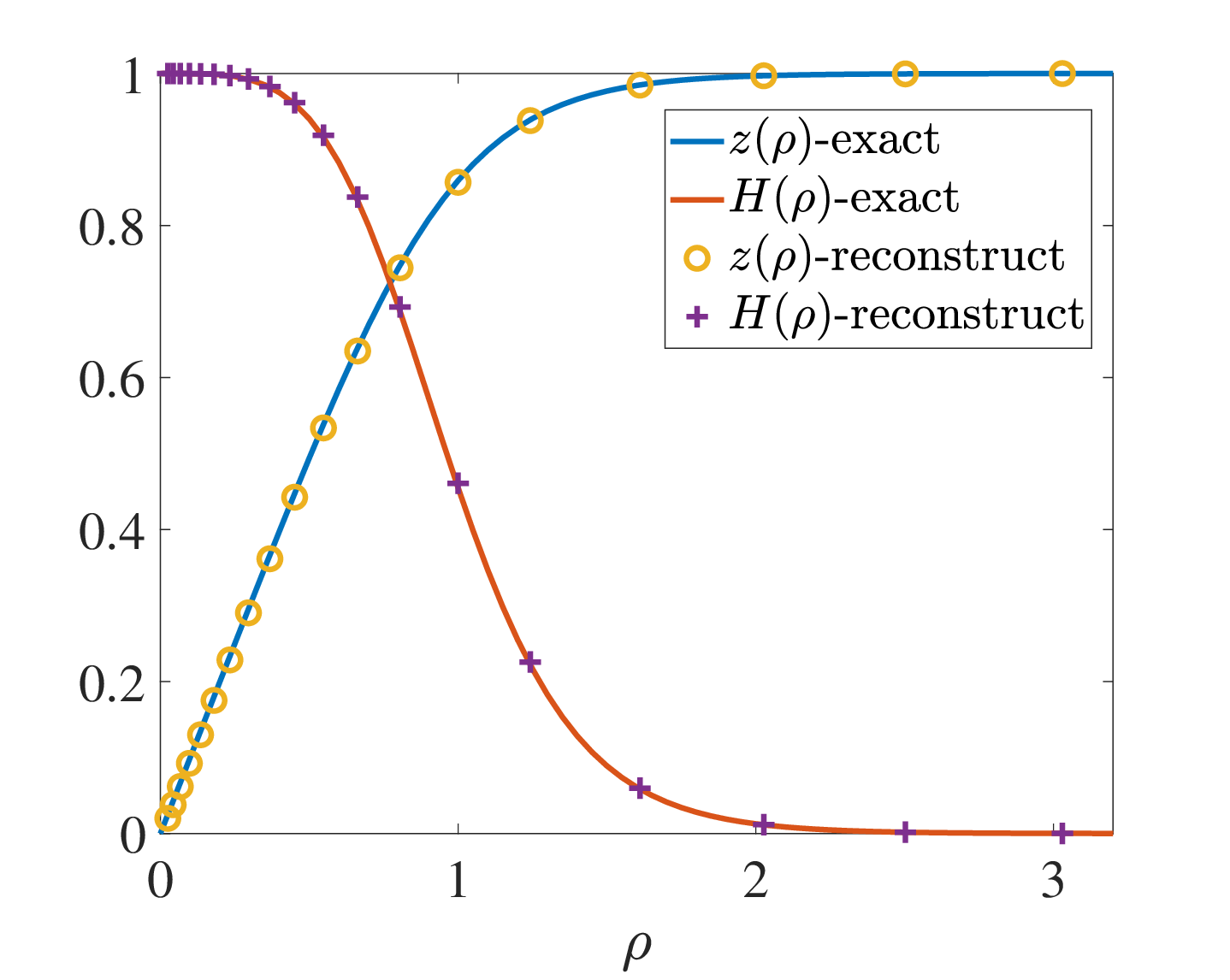}
      \caption{Comparison between the reconstructed metric components and their exact values reconstructed with the gauge field in BTZ black hole (left panel) and Schwarzschild-AdS black hole (right panel). We choose $k_1=0$ and $k_2=1$ for both two cases. }
      \label{scbtznp}
\end{figure}

For a BTZ black hole, the conductivity is given by \eqref{sigkw}. In fact, it is easy to see that this situation is essentially the same as the scalar field case, with a similar  method for solving inverse problem. The main difference is that, the gauge field has zero mass, and its conformal dimension is fixed. This means we cannot use conductivities with different conformal dimensions for our calculations. The reconstructed metric components and their exact values are shown in the left panel of Fig.~\ref{scbtznp}, here we choose $k_1=0$ and $k_2=1$. We see that the reconstructed method gives excellent results.

\subsubsection*{Schwarzschild-AdS black hole}
Similar to the case of the BTZ black hole, after setting $k_1=0$ and $k_2=1$, we can calculate two conductivities with these $k$. We also reconstruct the metric of the 5-dimensional Schwarzschild-AdS black hole using these two conductivities. Use \eqref{mathcalH} to obtain $h(\rho)$ and take $h(\rho)$ and $V_{k_1}$ or $V_{k_2}$ back into \eqref{mxwzkV} to reconstruct $z(\rho)$. The comparison of the reconstructed results of our methods and the exact results is shown in right panel of Fig.~\ref{scbtznp}. As demonstrated, we find that the reconstructed results agree well with the expected ones. Both this example and the BTZ black hole case show that our method is effective.

%

\subsection{Bulk reconstruction of charged background}
In the previous section, we discussed how to compute the two-point correlation functions of gauge fields, the relationship between these correlation functions and the holographic optical conductivity, and utilized the conductivity to reconstruct the bulk geometry in both BTZ black hole and five-dimensional Schwarzschild-AdS black hole. In fact, the similar method can be used to reconstruct all static planar/spherically/hyperbolically symmetric neutral black hole of which the bulk dimension is not four. In this subsection we will address the reconstruction problem of one charged background with nonzero static electric field.

The action of the gravity theory which describes a charged black hole and scalar fields in asymptotically AdS spacetime is given by
\begin{equation}\label{vecact11}
	S=\int \td^{d+1}x \sqrt{-g}\left[R+\frac{d(d-1)}{L^2}-\frac{1}{4}F_{\mu\nu}F^{\mu\nu}\right],
\end{equation}
here $F_{\mu\nu}$ is the Maxwell tensor, and $R$ is the Ricci scalar. Let us consider the $5$-dimensional charged black hole, and its metric ansatz can be written as
\begin{equation}\label{RNADSme}
     \td s^2=\frac{1}{z^2(\rho)}\left[h(\rho)(-\td t^2+\td \rho^2)+ \td x^2 + \td y^2 + \td u^2\right].
\end{equation}
Similarly, for the gauge field $A$, we still adopt the ansatz:
\begin{equation}
    A=\phi(\rho)\td t\,.
\end{equation}
To consider the direct and inverse of problem of gauge field correlation function in this case, we still add the gauge perturbation $A_{x}=z^{1/2}\mathcal{A}_{x}(\rho)\te^{\text{i}\omega t-\text{i}ky}$. However, in the case that $t-$component of background gauge potential is nonzero, we have to open the perturbations of metric meanwhile so that Maxwell equation and Einstein equation keep self-consistent in the linear order of perturbations. The self-consistent perturbed ansatz on metric reads $g_{tx}=z^{1/2}\tilde{g}_{tx}\te^{\text{i}\omega t-\text{i}ky}$ and $g_{xy}=z^{1/2}\tilde{g}_{xy}\te^{\text{i}\omega t-\text{i}ky}$. We obtain the equations of motion for the gauge field as follows:
\begin{equation}\label{mxwsc}
     -\frac{\td^2\mathcal{A}_{x}}{\td\rho^2}+\left(V_{k}+\frac{3}{4\rho^2}\right)\mathcal{A}_{x} +\frac{k}{\omega}\left(z   {g'_{xy}} +\frac{3 z' {g_{xy}} }{2}\right)\phi'=\omega^{2}\mathcal{A}_x,
\end{equation}
here
\begin{equation}\label{mxwsce}
\begin{split}
  V_{k}&=-\frac{z''}{2z}+\frac{3}{4}\left(\frac{z'^2}{z^2}-\frac1{\rho^2}\right)+k^2h+\frac{4 z^{2} \phi'^{2}}{h}.
  \end{split}
\end{equation}
If $\phi'=0$, the inhomogeneous terms of Eq.~\eqref{mxwsc} vanish entirely, reducing \eqref{mxwsc} to the previously discussed cases of the BTZ and Schwarzschild-AdS black holes in the earlier section. However, when the background is a charged black hole, of which the $t$-component of guage field $\phi'$ is nonzero, we see that inhomogeneous terms involving $g_{xy}$ and $g'_{xy}$ now appear in \eqref{mxwsc}.

One may wonder that if we could generalize our method of solving inverse scattering problem so that it could work for the equation containing inhomogeneous terms. However, solving for $z(\rho)$ and $h(\rho)$ now requires equation of motion of $g_{xy}$, which is interdependent with $g_{tx}$. Their coupled equations of motion are as follows:
\begin{equation}\label{gxy}
    \frac{\td^{2}}{\td \rho^{2}}{g_{xy}}=\left(-\frac{2 z^{2} \phi'^{2}}{h}-\omega^{2}+\frac{h''}{h}-\frac{h'^2}{h^{2}}-\frac{15 z''  }{2 z }+\frac{63z'^2}{4 z^2}-\frac{12 h}{z ^{2}}\right) {g_{xy}} -\frac{{g_{tx}}   k \omega}{z  ^{\frac{3}{2}}},
\end{equation}
\begin{equation}\label{gtx}
    \frac{\td}{\td \rho}{g_{tx}} =-\frac{ k h}{\omega}\left(z  ^{\frac{3}{2}} g'_{xy}+\frac{3 \sqrt{z }z' g_{xy}  }{2}\right)-4 z^{\frac{5}{2}} \phi'\mathcal{A}_x.
\end{equation}
And the equation of motion of $\phi$ is given by
\begin{equation}\label{phieom}
    \frac{\td^{2}}{\td\rho^{2}}\phi=\frac{h'z+hz'}{hz}\frac{\td}{\td\rho}\phi.
\end{equation}
By solving the simultaneous two potential $V_{k_1}$ and $V_{k_2}$ from \eqref{mxwsce} with $k_1$ and $k_2$, together with \eqref{gxy}, \eqref{gtx}, \eqref{phieom}, we cannot determine $z(\rho)$, $h(\rho)$, $g_{tx}(\rho)$, $g_{xy}(\rho)$ and $\phi(\rho)$ and $\mathcal{A}_x(\rho)$. We need also at last one more equation about those unknowns, for example, a correlation function of $g_{xy}$.
\begin{figure}
      \centering
      \includegraphics[width=0.6\textwidth]{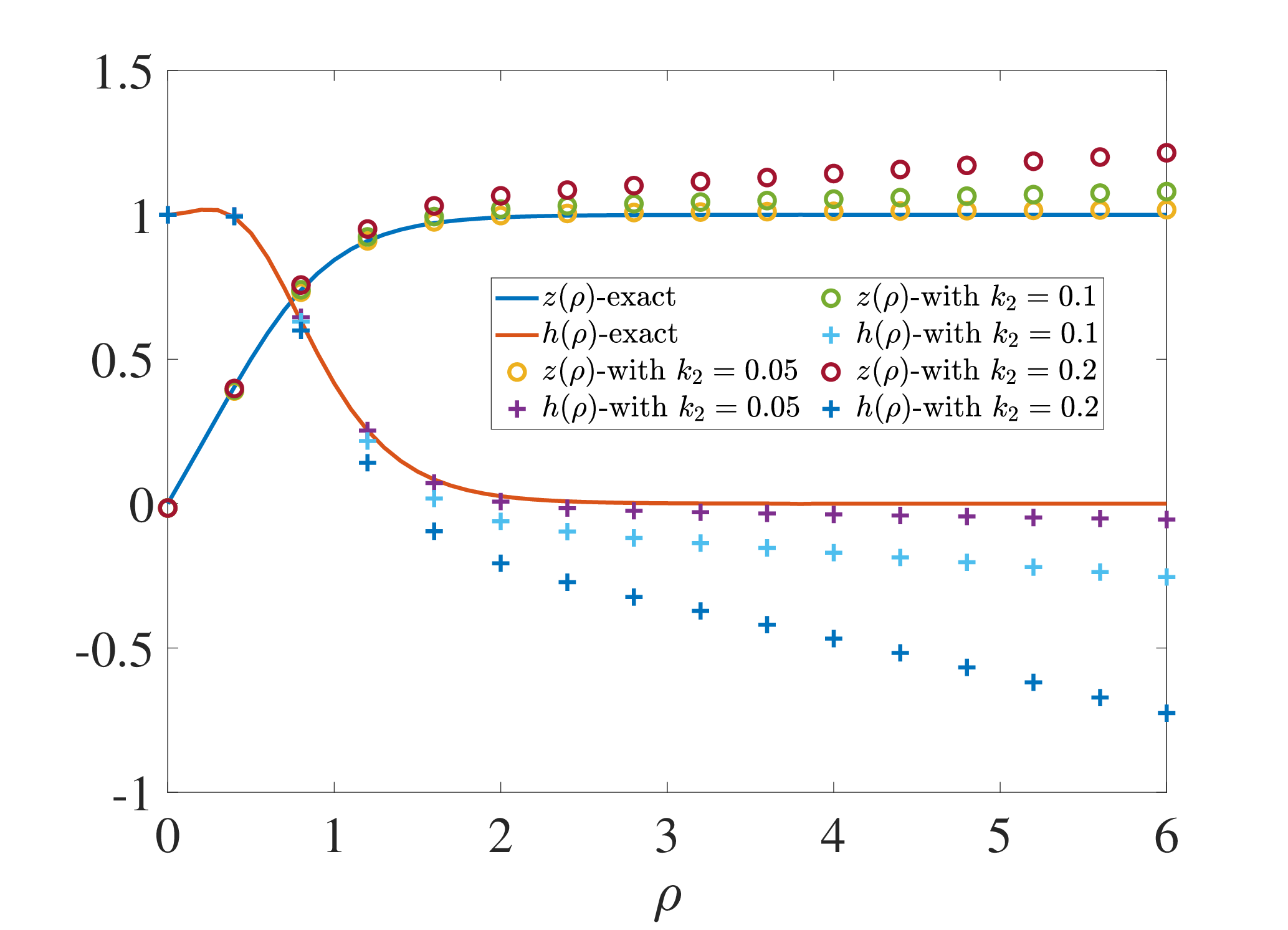}
      \caption{Reconstruction of metric in RN-AdS black hole by choosing $k_1 = 0$ and $k_2 = 0.05,\ 0.1,\ 0.2$. Here we choose $z_h=1$ and black hole charge $Q=Q_m/4$, where $Q_m$ is the charge for extremal RN-AdS black hole. }

      \label{RNADSoio}
\end{figure}
It needs us to have some prior information of the gravitational theory. This introduces a circular argument: the reconstruction process assumes aspects of the gravitational theory it aims to determine.

The problem may be solved partially by following approximation way. We note that the inhomogeneous term in \eqref{mxwsc} can be approximated by adopting a reconstruction approach that makes use of a small $k$. Since we need at least two correlation functions with two different $k$ values to reconstruct the metric, when one of the $k$ values is chosen to be zero, the equation always holds, and another small $k$ can be selected for the reconstruction.

As an example, let us consider the Reissner-Nordstr\"{o}m-AdS (RN-AdS) black hole. We can set $h(\rho)$ and $z(\rho)$ in \eqref{RNADSme} as following: $h(\rho)=1+z^2(\rho)-(Q^2+2)z^3(\rho)+Q^2z^4(\rho)$, and the relationship between $ \rho $ and $ z (\rho)$ is similarly described by the following algebraic equation:
\begin{equation}\label{rnadsmr}
    \rho=\sum_{1+\alpha^2-(Q^2+2)\alpha^3+Q^2\alpha^4=0}{\left[\frac{\ln(z-\alpha)}{4Q^2\alpha^3-3(Q^2+2)\alpha^2+2\alpha}-{\frac{\ln(-\alpha)}{4Q^2\alpha^3-3(Q^2+2)\alpha^2+2\alpha}}\right]}.
\end{equation}
Here $Q$ is the charge of the black hole\footnote{Here $Q^2=4$ represents an extremal black hole.}. We perform the reconstruction by choosing $k_1 = 0$ and $k_2 = 0.05,\ 0.1,\ 0.2$, respectively. Fig. \ref{RNADSoio} shows the results reconstructed with $k_2 = 0.05,\ 0.1,\ 0.2$. For the case that $k_2$ is small, we see that reconstruction gives a well-approximated bulk metric. For large $k_2$, it can be seen that the reconstructed results agree well with the exact solution at small values of $\rho$, but the discrepancy grows larger as $\rho$ increases. This error mainly arises because, at small $\omega$, the ratio of $k$ to $\omega$ becomes large, which means the inhomogeneous term has a relatively significant impact and so the $k/\omega$ in Eq.~\eqref{mxwsc} cannot be neglected. Since errors at small $\omega$ primarily affect the behavior at large $\rho$, the reconstruction error also becomes more pronounced in the large-$\rho$ region. This type of error originates from the intrinsic properties of the field's equation of motion, and our computational method also contributes to it. We will provide a detailed explanation in the next section.

\section{Stability of the method}
\begin{figure}

      \includegraphics[width=0.49\textwidth]{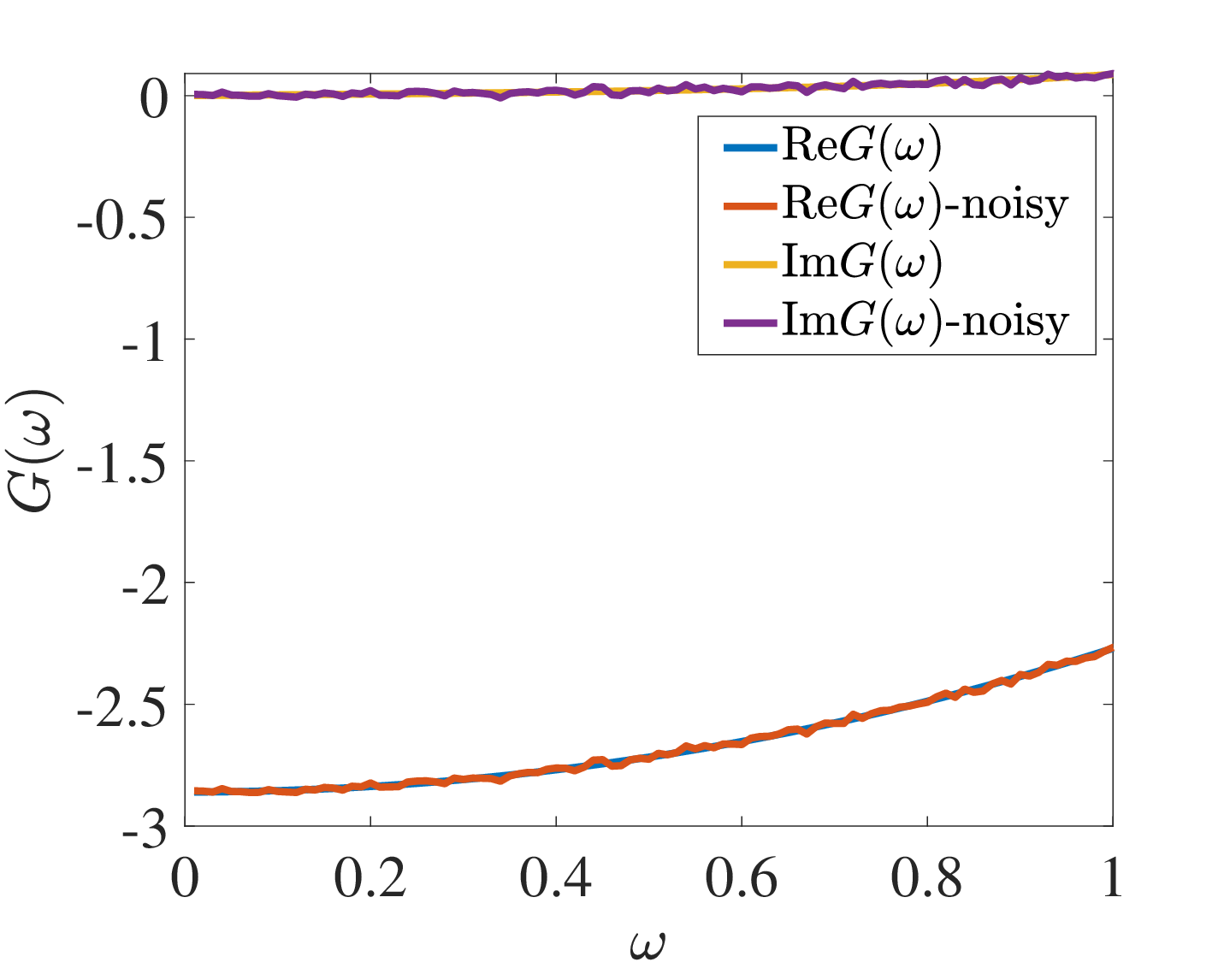}
      \includegraphics[width=0.49\textwidth]{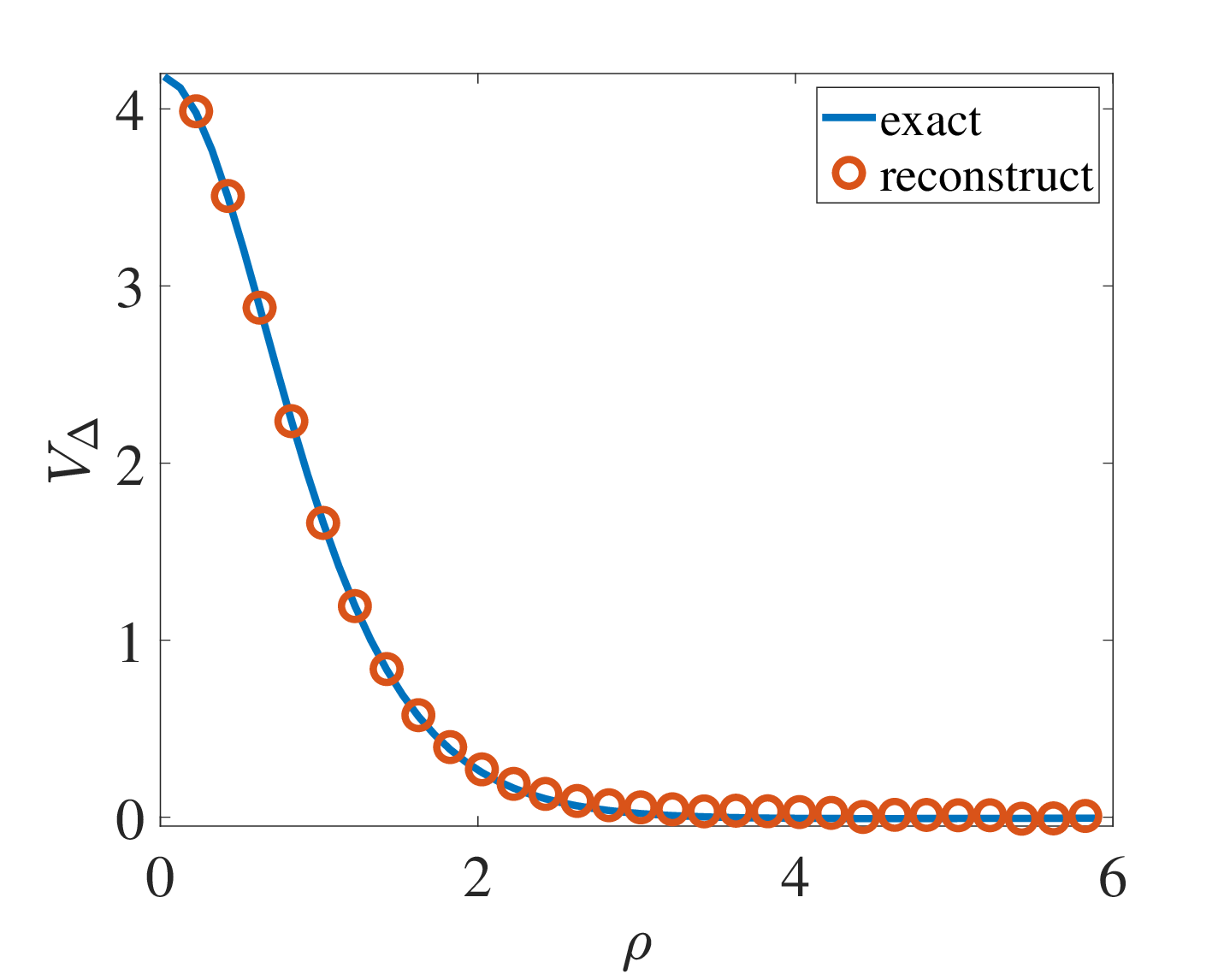}
      \caption{The perturbed two-point correlation function of a scalar in BTZ black hole (left panel) and the potential reconstructed from it (right panel). Here, $\sigma = 0.01$ is used, and it can be observed that the reconstruction results are relatively good.}
      \label{pec1}
\end{figure}
\begin{figure}

      \includegraphics[width=0.49\textwidth]{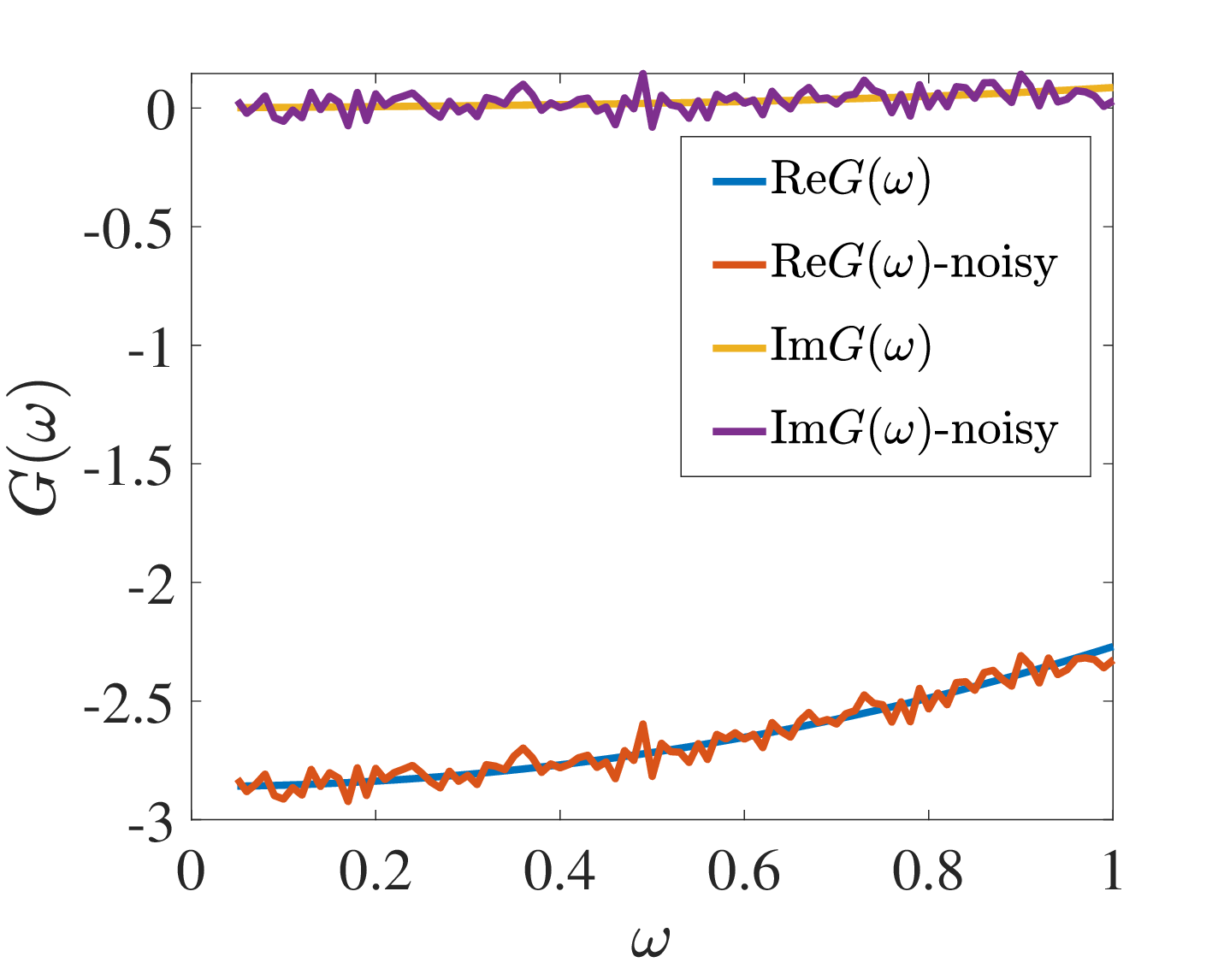}
      \includegraphics[width=0.49\textwidth]{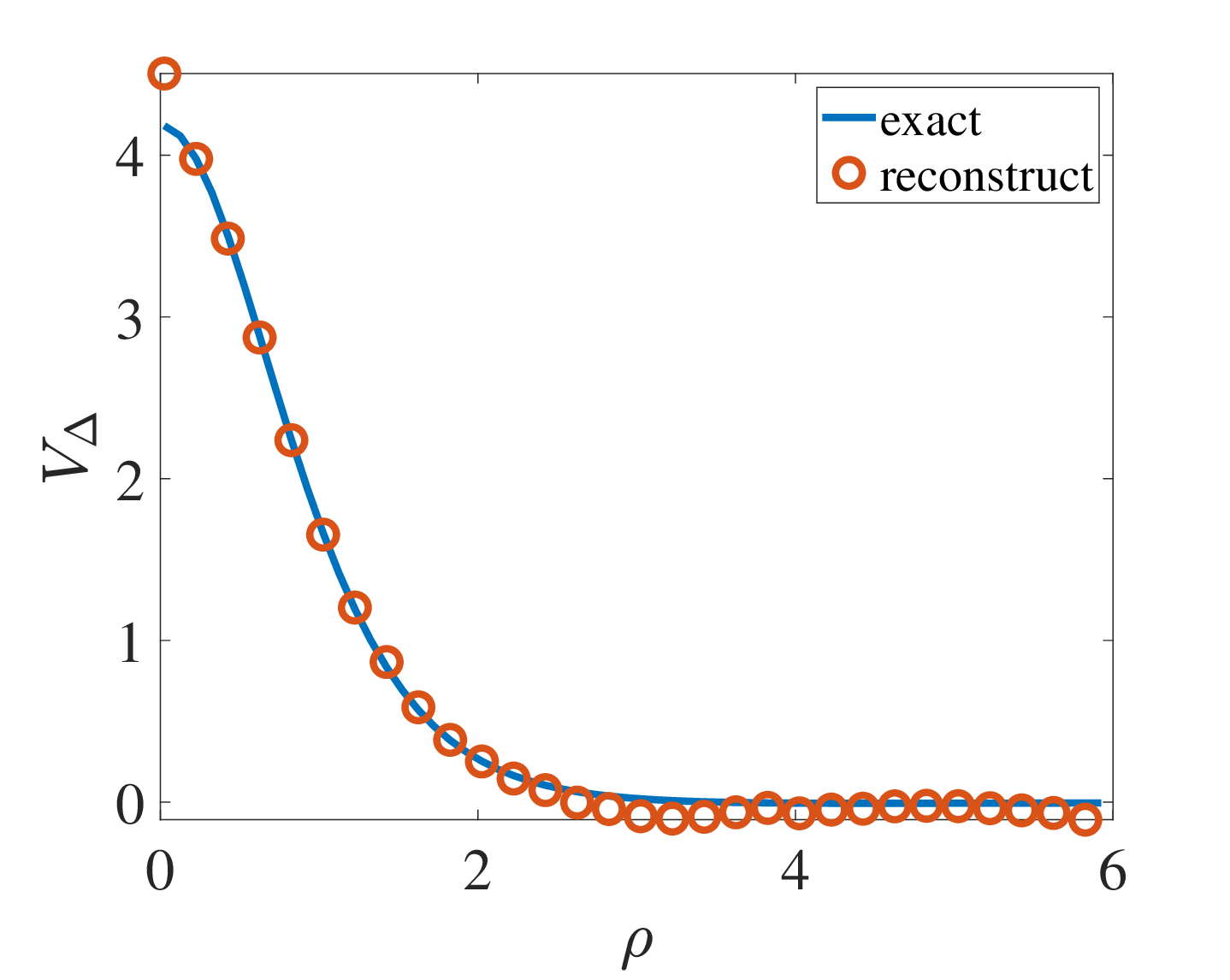}
      \caption{The perturbed two-point correlation function of a scalar in BTZ black hole (left panel) and the potential reconstructed from it (right panel). Here, $\sigma = 0.05$ is used. It can be seen that significant errors have already occurred when $\rho$ becomes large. }
      \label{pec5}
\end{figure}
\begin{figure}

      \includegraphics[width=0.49\textwidth]{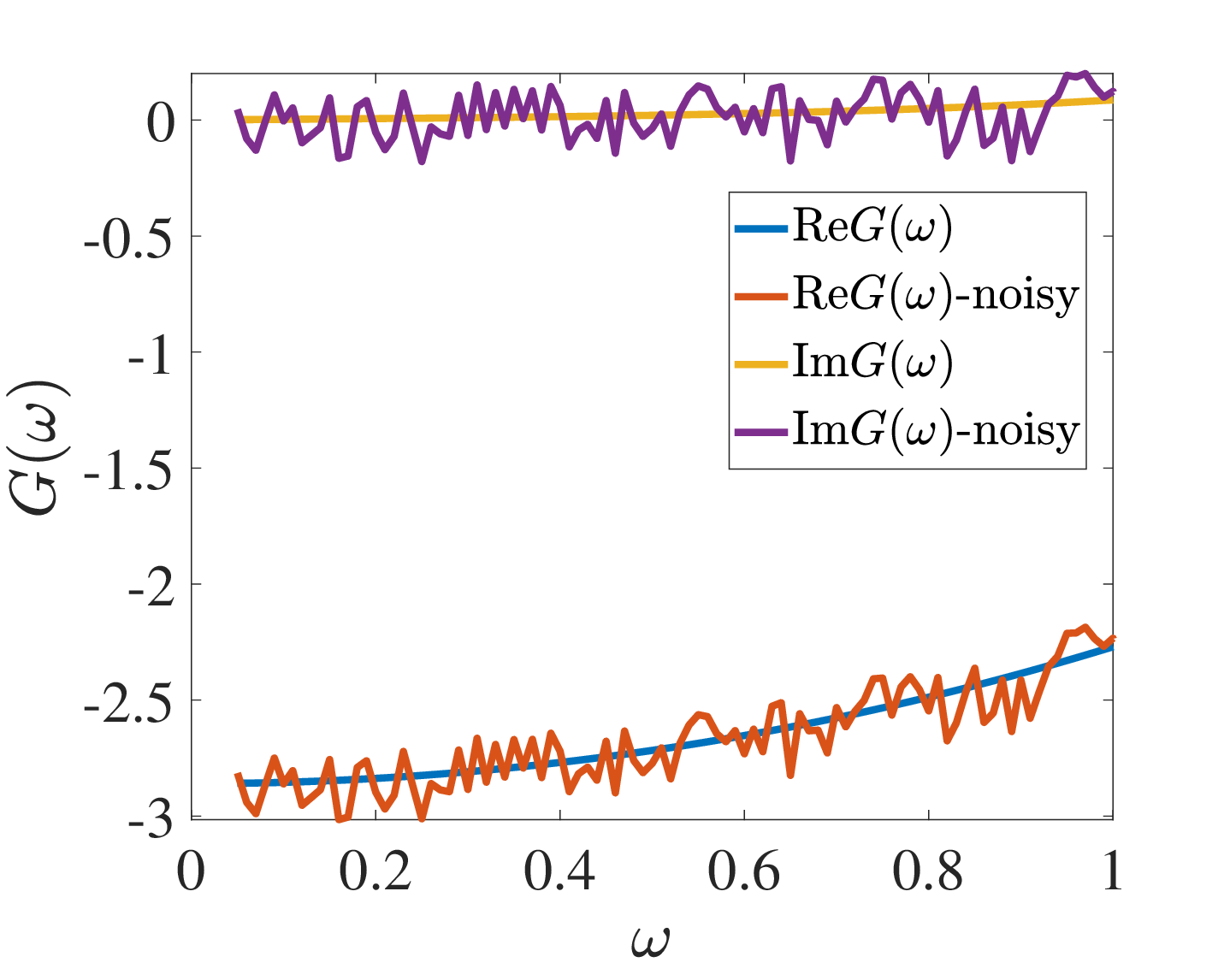}
      \includegraphics[width=0.49\textwidth]{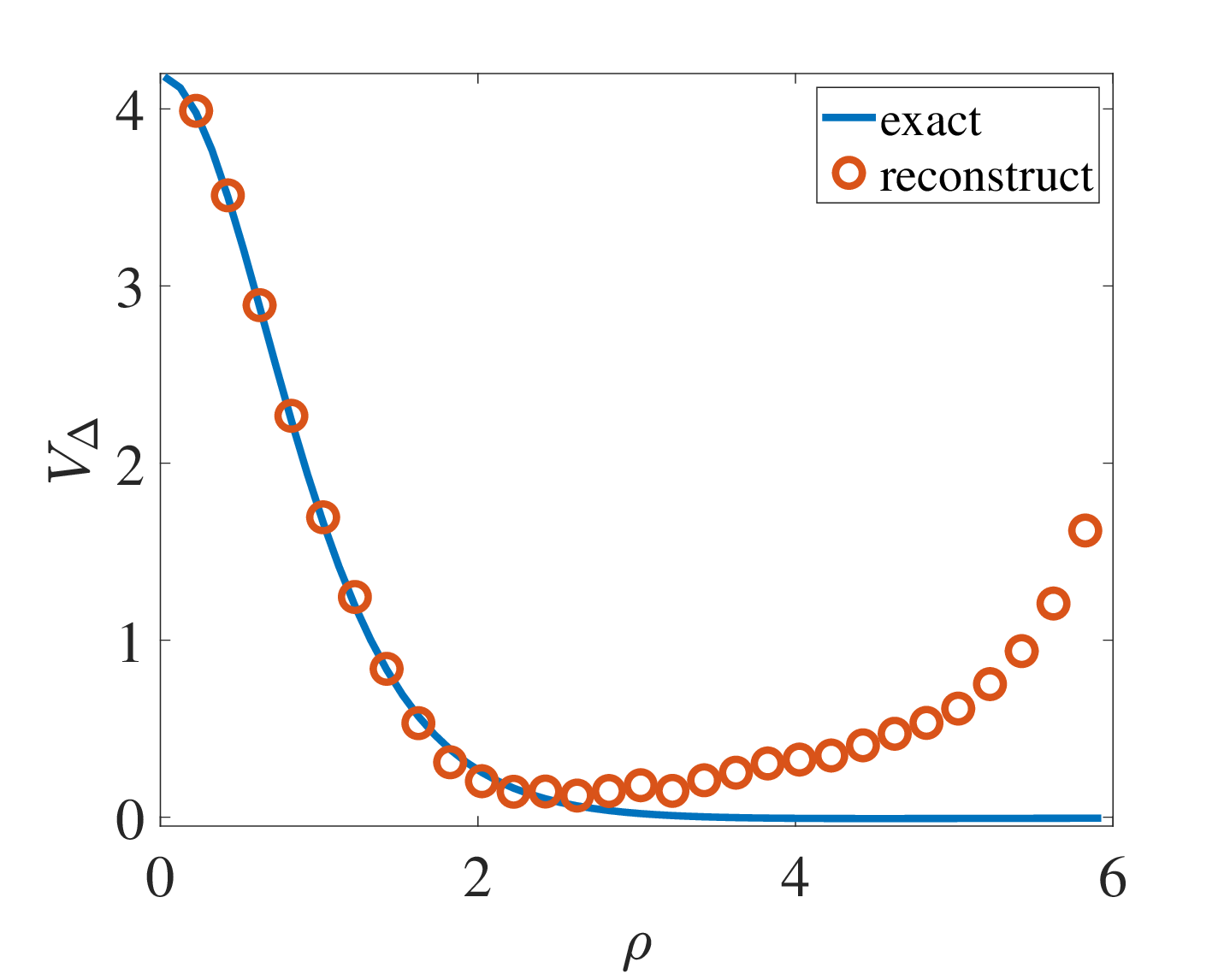}
      \caption{The perturbed two-point correlation function of a scalar in BTZ black hole (left panel) and the potential reconstructed from it (right panel). Here, $\sigma = 0.1$ is used. It can be seen that for larger $\rho$, the reconstruction results depart from the exact value dramatically. }
      \label{pec10}
\end{figure}

In the models described above, we employed the two-point correlation functions to reconstruct the metric. In experiments of real-world, the measured data are often subject to errors and noise, which raises a critical question: how stable is our method when experimental data are noisy? To evaluate this, we introduced perturbations to the exact correlation function using a Gaussian noise distribution $ N(\mu, \sigma^2) $. These perturbations were then incorporated into our computational process. Specifically, we defined the perturbed function as $ G'(\omega) = G(\omega) + g(\omega) $, where $ g(\omega) $ represents the noise in the experiments. Additionally, in experiments, we cannot measure data at $\omega\rightarrow\infty$, and we will also check influences caused by truncations.

In real experiments, noise often stems from systematic errors, which are independent of the signal's strength. This characteristic informed our choice to add noise in this manner. As an example, we consider the correlation fuction of a scalar in BTZ black hole. By setting $ \mu = 0 $ and varying $ \sigma $, we generated multiple function plots. Figures \ref{pec1}, \ref{pec5}, and \ref{pec10} illustrate the reconstructed potentials under different noise intensities. We tested several values of $ \sigma $ ranging from 0.01 to 0.5. For each case, we calculated the standard deviation between the reconstructed results and the exact values of effective potential.

In real experimental scenarios, low-frequency measurements are generally more accurate, while high-frequency measurements are more prone to error. Consequently, the reconstructed potential and, by extension, the derived metric components to exhibit greater inaccuracies at smaller $ \rho $ values. At large $ \omega$ or $k$, the correlation function is governed by geometry of AdS boundary and so independent of the bulk metric so we have $G(\omega,k)=G_{\text{AdS}}(\omega,k)+\cdots$ for large enough $\omega$ or $k$. Here $G_{\text{AdS}}(\omega,k)$ is the correlation function in pure AdS spacetime. Understanding this pattern offers a practical advantage: it allows us to assess the reliability of experimental data. If necessary, truncation methods can be employed to minimize the impact of high-frequency noise, thereby improving reconstruction accuracy. We may use the function $G_{\text{AdS}}(\omega,k)$ to replace experimental data for large $\omega$ or $k$ if the material has been assumed to have holographic dual.

\begin{figure}

      \includegraphics[width=0.49\textwidth]{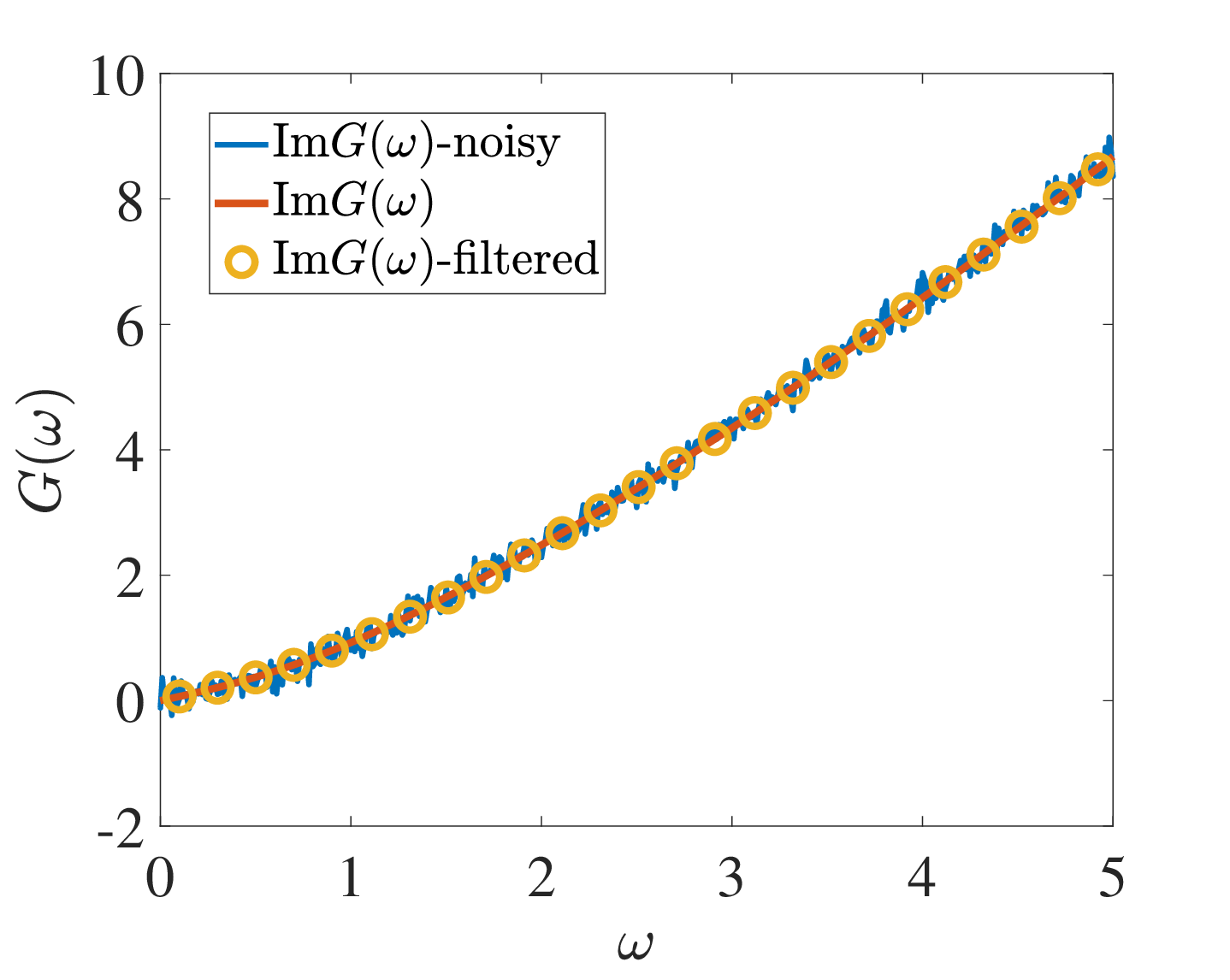}
      \includegraphics[width=0.49\textwidth]{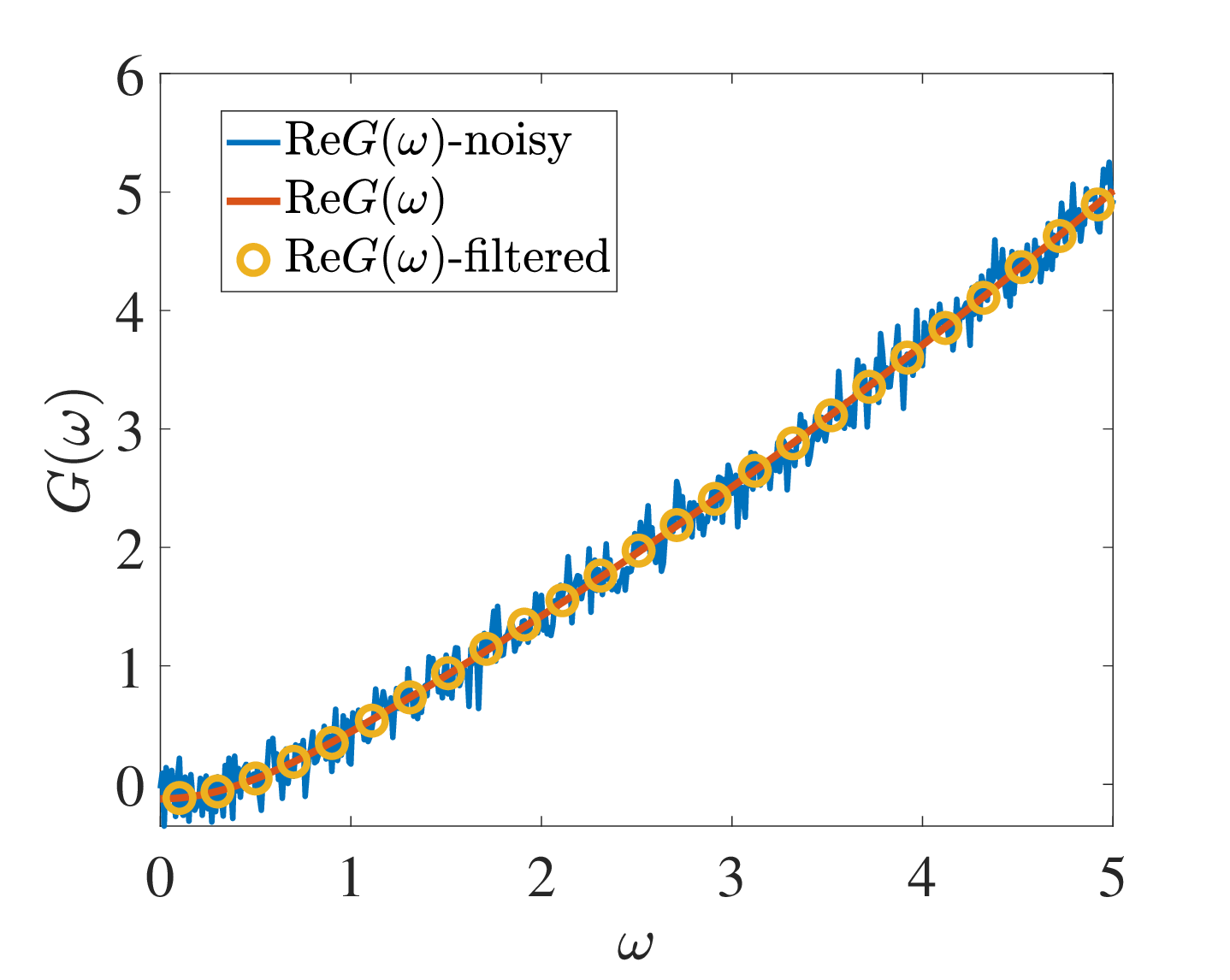}
      \caption{A comparison of the filtered correlation function of a scalar in BTZ black hole with the exact values and the experimentally measured noisy correlation function is shown in the figure. Here, $\sigma=0.15$ for the noise, $k=0$, $\nu=3/2$, and the filtering operation was performed 50 times. The left panel shows the imaginary part, while the right panel shows the real part.}
      \label{imagNreal}
\end{figure}
Let us now explain how to suppress the random noise in our reconstruction. In fact, our method utilizes only the imaginary part of the correlation function, while both the real and imaginary parts can be measured in experiments. There is a way to filter out noise using the real part of the correlation function. Mathematically, a relation known as the Kramers-Kronig relations connects the real and imaginary parts of any analytic complex function in the upper half-plane. These relations are often employed to compute the real part from the imaginary part of a response function (and vice versa) in physical systems. For stable systems, causality implies analyticity, and conversely, analyticity reflects the causality of a corresponding stable physical system. Through this relationship, the real and imaginary parts of an analytic complex function in the upper half-plane obey a Hilbert transform.

\begin{figure}
      \centering
      \includegraphics[width=0.55\textwidth]{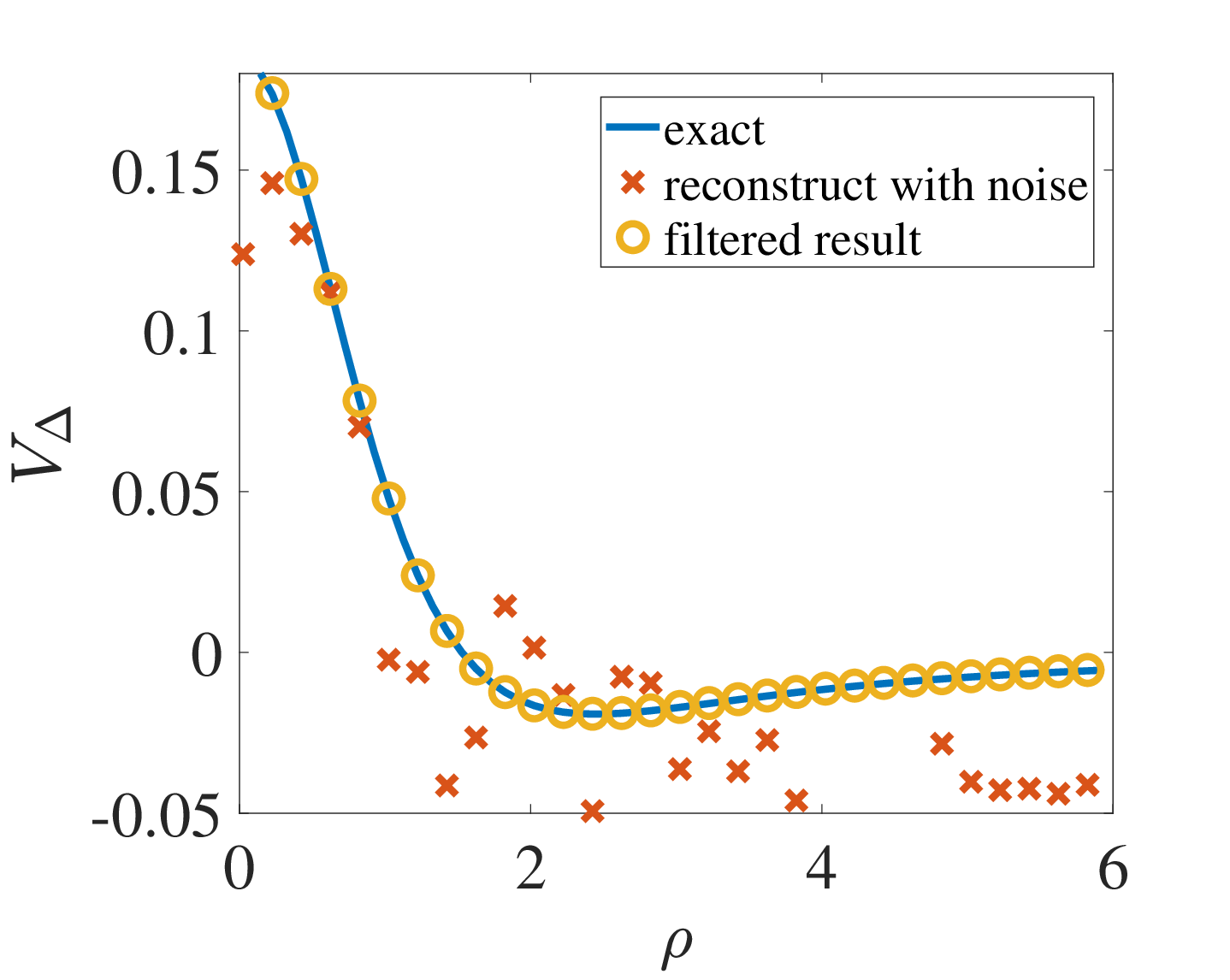}
      \caption{The comparison between the bulk reconstructed using the noisy measurement results directly and using the results after 50 filtering iterations is shown alongside the exact results. It is evident that the filtered results achieve significantly better accuracy. Here, we use the correlation function of a scalar in BTZ black hole and $k=0$, $\nu=3/2$ and we take $\sigma = 0.15$ for the noise.}
      \label{Verf}
\end{figure}

Consider a complex function $ \chi(\omega) $. Given the imaginary part $ \text{Im}[\chi(\omega)] $, the real part $ \text{Re}[\chi(\omega)] $ can be determined, and vice versa, using the Hilbert transform:

\begin{equation}
        \Re[\chi(\omega)] = \frac{1}{\pi} \mathcal{P} \int_{-\infty}^\infty \frac{\Im[\chi(\omega')]}{\omega' - \omega} \, d\omega',\
        \Im[\chi(\omega)] = -\frac{1}{\pi} \mathcal{P} \int_{-\infty}^\infty \frac{\Re[\chi(\omega')]}{\omega' - \omega} \, d\omega'.
\end{equation}
where $\mathcal{P} $ denotes the Cauchy principal value of the integral. This relationship allows us to obtain the corresponding imaginary part $\Im(\mathcal{G})_1$ from the measured real part $\Re(\mathcal{G})$ of the correlation function, and the corresponding real part $\Re(\mathcal{G})_1$ from the measured imaginary part $\Im(\mathcal{G})$ of the correlation function. Then we take the average of $\Im(\mathcal{G})_1$ and $\Im(\mathcal{G})$ to obtain $\Im(\mathcal{G})_1'$, and the average of $\Re(\mathcal{G})_1$ and $\Re(\mathcal{G})$ to obtain $\Re(\mathcal{G})_1'$. We then obtain a new correlation function $\mathcal{G}_1'$ and use it to obtain $\mathcal{G}_2'$ by similar steps. By repeating the above steps, we can obtain the filtered correlation function $\mathcal{G}_n'$, which can then be used for our reconstruction calculations. The real and imaginary parts after filtering 50 times are shown in Fig.~\ref{imagNreal}. We see that the noise is almost removed from the data. Moreover, the reconstruction results after using this filtering method are significantly better than those obtained through direct calculation, as shown in Fig.~\ref{Verf}. It can be seen that before applying the filter, the effective potential reconstructed directly from the noisy measured two-point correlation function deviates significantly from the exact effective potential, with large errors. However, after applying the filter, the reconstructed effective potential shows much better agreement with the exact result. This indicates that the filtering process effectively reduces many of the errors introduced by the measurement noise.

\section{Summary}
In this work, we developed a unified and versatile framework for holographic bulk reconstruction based on solving inverse scattering problems, leveraging boundary two-point correlation functions. This approach allows one to extract bulk geometric information using either scalar or gauge field probes and is particularly well-suited to static asymptotically AdS spacetimes with planar, spherical, or hyperbolic symmetries.

A central advancement of our method lies in the utilization of momentum dependence, which enables bulk reconstruction using only a single operator-eliminating the need for multiple operators with different conformal dimensions. This makes the method more practical for experimental applications and reduces potential systematic errors introduced by multiple field insertions. For gauge fields, which inherently lack mass and thus have fixed conformal dimensions, momentum plays a similarly critical role, allowing metric reconstruction through conductivities measured at distinct momenta. We successfully demonstrated our method across several holographic backgrounds, including the BTZ black hole, Schwarzschild-AdS, and RN-AdS geometries. These examples illustrate the broad applicability of our framework and its robustness even in the presence of coupling between fields or metric backreaction effects-though the latter still presents challenges in fully coupled charged systems. In such cases, naive application of inverse methods may lead to overdetermined systems or require prior knowledge of the gravitational theory, potentially introducing circular reasoning.

The ability to independently reconstruct electromagnetic and scalar-probed metrics holds profound implications beyond theoretical studies, particularly in experimental and material science applications. Consider a material suspected of exhibiting holographic properties: by probing it with distinct external sources, such as electric fields, magnetic fields or scalar-coupled stimuli, researchers could measure unique response functions, each enabling the reconstruction of a corresponding effective metric. Consistency across these reconstructed geometries would strongly indicate holographic behavior, as the material's responses mirror a unified bulk structure. Conversely, metric discrepancies would challenge holographic interpretations, pointing instead to emergent phenomena driven by competing couplings or unresolved internal degrees of freedom. This framework gains further relevance in systems like candidate holographic superconductors. For materials displaying metric consistency, detailed temperature-dependent conductivity measurements could map how the reconstructed geometry evolves thermally. Such correlations might unveil predictive relationships, for instance, inferring critical behavior near phase transitions or extreme-condition properties that resist direct measurement.

Another result in this work is the stability analysis under noisy or incomplete data. We introduced a filtering technique based on the Kramers-Kronig relations to iteratively improve the quality of the measured correlation functions. This approach effectively mitigates noise and enhances the reliability of the reconstructed metric, especially in regions where the data are most sensitive to errors (e.g., large-$\rho$ or near-horizon regions). This suggests that our method is not only theoretically sound but also experimentally viable under realistic measurement conditions.

Nonetheless, several open problems remain. Our analysis assumes static backgrounds and does not extend to rotating or dynamical geometries, which would require significant generalization of the current formalism. Additionally, conformal invariance in 4-dimensional Maxwell theory limits the reconstructible metric components, emphasizing the need for hybrid methods that combine different types of probes. For the case that background is charged, our method can be used to reconstruct metric approximately by using correlation functions of small momentums, so it is still interesting to find an exact method for solving inverse problem in such case.

\begin{acknowledgments}
This work is supported by the Natural Science Foundation of China under Grant No. 12375051 and Tianjin University Self-Innovation Fund Extreme Basic Research Project Grant No. 2025XJ22-0014 and 2025XJ21-0007.
\end{acknowledgments}

\bibliographystyle{JHEP}
\bibliography{ahr}
\end{document}